\documentclass[reprint,amsmath,amssymb,aps,prd,]{revtex4-1}
\usepackage{mathrsfs}
\usepackage{graphicx}
\usepackage{epsfig,graphics,subfigure,psfrag,amsmath,amssymb}
\usepackage{tabularx}
\newcolumntype{C}{>{\centering\arraybackslash}X}
\usepackage{dcolumn}
\usepackage{bm}
\usepackage{overpic}
\usepackage{xspace}
\usepackage{rotating}
\usepackage{color}
\usepackage{multirow}
\usepackage{colortbl}
\usepackage{epstopdf}
\usepackage{float}
\usepackage{lineno}
\usepackage{placeins}

\definecolor{aa}{RGB}{0,0,255}
\usepackage[bookmarksnumbered=true,colorlinks,urlcolor=aa,linkcolor=blue,anchorcolor=aa,citecolor=aa]{hyperref}

\DeclareRobustCommand{\PreserveBackslash}[1]{\let\temp=\\#1\let\\=\temp}
\newcolumntype{C}[1]{>{\PreserveBackslash\centering}p{#1}}
\newcolumntype{R}[1]{>{\PreserveBackslash\raggedleft}p{#1}}
\newcolumntype{L}[1]{>{\PreserveBackslash\raggedright}p{#1}}

\DeclareRobustCommand{\g}{\gamma}

\DeclareRobustCommand{\pp}{\pi^+\pi^-}
\DeclareRobustCommand{\piz}{\pi^{0}}

\DeclareRobustCommand{\jpsi}{J/\psi}
\DeclareRobustCommand{\psip}{\psi(2S)}

\DeclareRobustCommand{\ra}{\rightarrow}
\DeclareRobustCommand{\gevcc}{{\rm GeV}/c^2}
\DeclareRobustCommand{\gev}{\mathrm{GeV}}

\DeclareRobustCommand{\mevcc}{{\rm MeV}/c^2}

\DeclareRobustCommand{\ee}{e^{+}e^{-}}
\DeclareRobustCommand{\DszDszb}{D^{*0}\Bar{D}^{*0}}
\DeclareRobustCommand{\Dz}{D^{0}}
\DeclareRobustCommand{\Dzb}{\Bar{D}^{0}}
\DeclareRobustCommand{\Dsz}{D^{*0}}
\DeclareRobustCommand{\Dszb}{\Bar{D}^{*0}}
\DeclareRobustCommand{\ks}{K_{S}^{0}}
\DeclareRobustCommand{\dpizDz}{D^{*0} \rightarrow \pi^{0} D^{0}}
\DeclareRobustCommand{\dgamDz}{D^{*0} \rightarrow \gamma D^{0}}
\DeclareRobustCommand{\ppDD}{D^{*0}\bar{D}^{*0} \rightarrow \pi^{0}\pi^{0}D^{0}\bar{D}^{0}}
\DeclareRobustCommand{\gampiDD}{D^{*0}\bar{D}^{*0} \rightarrow \gamma\pi^{0}D^{0}\bar{D}^{0}}
\DeclareRobustCommand{\x}{X(4013)}
\DeclareRobustCommand{\etac}{\eta_{c}(3S)}
\DeclareRobustCommand{\chiczp}{\chi_{c0}(3P)}
\DeclareRobustCommand{\chicop}{\chi_{c1}(3P)}
\DeclareRobustCommand{\chictp}{\chi_{c2}(3P)}
\DeclareRobustCommand{\chicjp}{\chi_{c0,1,2}(3P)}
\DeclareRobustCommand{\dx}{X(4013) \rightarrow D^{*0}\Bar{D}^{*0}}
\DeclareRobustCommand{\detac}{\eta_{c}(3S) \rightarrow D^{*0}\Bar{D}^{*0}}
\DeclareRobustCommand{\dchiczp}{\chi_{c0}(3P) \rightarrow D^{*0}\Bar{D}^{*0}}
\DeclareRobustCommand{\dchicop}{\chi_{c1}(3P) \rightarrow D^{*0}\Bar{D}^{*0}}
\DeclareRobustCommand{\dchictp}{\chi_{c2}(3P) \rightarrow D^{*0}\Bar{D}^{*0}}

\DeclareRobustCommand{\konepi}{\Dz\ra K^{-}\pi^{+}}
\DeclareRobustCommand{\kthreepi}{\Dz\ra K^{-}\pi^{+}\pi^{+}\pi^{-}}
\DeclareRobustCommand{\kspipi}{\Dz\ra K_{S}^{0}\pi^{+}\pi^{-}}

\DeclareRobustCommand{\kmpip}{K^{-}\pi^{+}}

\DeclareRobustCommand{\kmpipipi}{K^{-}\pi^{+}\pi^{+}\pi^{-}}

\DeclareRobustCommand{\kszpipi}{K_{S}^{0}\pi^{+}\pi^{-}}

\def\gev   {\ensuremath{\mbox{\,GeV}\xspace}}

\def\babar{\mbox{\slshape B\kern-0.1em A\kern-0.1em B\kern-0.1em{A\kern-0.2em R}}}

\begin{document}
\graphicspath{{figure/}}
\DeclareGraphicsExtensions{.eps,.png,.ps}
\title{\boldmath Search for charmonium(like) states $X$ in $\ee\ra\g X \ra\g\DszDszb$ at BESIII}

\author{
  \begin{small}
    \begin{center}
M.~Ablikim$^{1}$\BESIIIorcid{0000-0002-3935-619X},
M.~N.~Achasov$^{4,c}$\BESIIIorcid{0000-0002-9400-8622},
P.~Adlarson$^{82}$\BESIIIorcid{0000-0001-6280-3851},
X.~C.~Ai$^{88}$\BESIIIorcid{0000-0003-3856-2415},
C.~S.~Akondi$^{31A,31B}$\BESIIIorcid{0000-0001-6303-5217},
R.~Aliberti$^{39}$\BESIIIorcid{0000-0003-3500-4012},
A.~Amoroso$^{81A,81C}$\BESIIIorcid{0000-0002-3095-8610},
Q.~An$^{78,65,\dagger}$,
Y.~H.~An$^{88}$\BESIIIorcid{0009-0008-3419-0849},
M.~S.~Anderson$^{39}$\BESIIIorcid{0009-0008-1550-2632}
% ...
\\
\vspace{0.2cm}
(BESIII Collaboration)\\
\vspace{0.2cm} {\it
$^{1}$ Institute of High Energy Physics, Beijing 100049, People's Republic of China\\
% ...
$^{4}$ Budker Institute of Nuclear Physics SB RAS (BINP), Novosibirsk 630090, Russia\\
% ...
$^{31}$ INFN Sezione di Ferrara, (A)INFN Sezione di Ferrara, I-44122, Ferrara, Italy; (B)University of Ferrara, I-44122, Ferrara, Italy\\
% ...
$^{39}$ Johannes Gutenberg University of Mainz, Johann-Joachim-Becher-Weg 45, D-55099 Mainz, Germany\\
% ...
$^{65}$ State Key Laboratory of Particle Detection and Electronics, Beijing 100049, Hefei 230026, People's Republic of China\\
% ...
$^{78}$ University of Science and Technology of China, Hefei 230026, People's Republic of China\\
% ...
$^{81}$ University of Turin and INFN, (A)University of Turin, I-10125, Turin, Italy; (B)University of Eastern Piedmont, I-15121, Alessandria, Italy; (C)INFN, I-10125, Turin, Italy\\
$^{82}$ Uppsala University, Box 516, SE-75120 Uppsala, Sweden\\
% ...
$^{88}$ Zhengzhou University, Zhengzhou 450001, People's Republic of China\\
\vspace{0.2cm}
$^{\dagger}$ Deceased\\
% ...
$^{c}$ Also at the Novosibirsk State University, Novosibirsk, 630090, Russia\\
% ...
}
\end{center}
    \vspace{0.4cm}
\end{small}
}
\affiliation{}


\begin{abstract}
A search is performed for a state $X$ decaying into $\Dsz\Dszb$ produced in the process $\ee\to\gamma X$ using a data sample corresponding to an integrated luminosity of 1667.4~$\rm pb^{-1}$ collected at $\sqrt{s} = 4.682$~GeV with the BESIII detector at the BEPCII. The state $X$ could be one of the $C$-even states $\x$, $\etac$, $\chiczp$, $\chicop$, or $\chictp$. No significant signal is observed in the corresponding signal region. Upper limits of $\sigma_{\ee\ra\gamma X}\cdot {\rm Br}_{X\ra\Dsz\Dszb}$ at 90\% confidence level are provided, where $\sigma_{\ee\ra\gamma X}$ represents the cross section of the $\ee\ra\gamma X$ process, and ${\rm Br}_{X\ra\Dsz\Dszb}$ is the branching fraction of the $X\ra\Dsz\Dszb$ process. 
\end{abstract}

\newcommand{\BESIIIorcid}[1]{\href{https://orcid.org/#1}{\hspace*{0.1em}\raisebox{-0.45ex}{\includegraphics[width=1em]{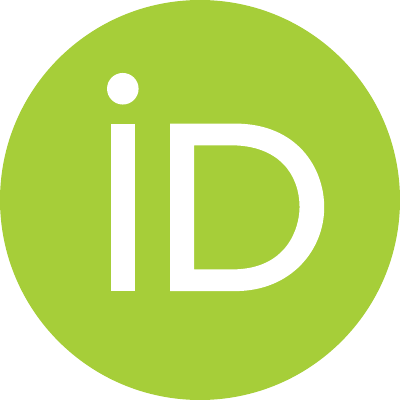}}}}

\maketitle
\section{INTRODUCTION}
The $X(3872)$ state (also known as $\chi_{c1}(3872)$) was first observed twenty years ago by Belle~\cite{intro:1-1} in $B^{\pm}\ra K^{\pm}\pi^{+}\pi^{-}\jpsi$ decays, and was subsequently confirmed by several other experiments~\cite{intro:1-2, intro:1-3, intro:1-4}. Since its discovery, the $X(3872)$ has stimulated considerable interest. $\babar$ observed the decays $X(3872)\ra\jpsi\gamma$ and $X(3872)\ra\psip\gamma$, and Belle also observed the decay $X(3872)\ra\jpsi\gamma$, which confirms that the $X(3872)$ is a $C$-even state~\cite{intro:1-5, intro:1-5-1, intro:1-6}. 
The CDF and LHCb experiments determined the spin-parity of the $X(3872)$ to be $J^{P}=1^{+}$~\cite{intro:1-7, intro:1-8}, and CDF found the $\rho^{0}(770)$ resonance dominating the $\pi^{+}\pi^{-}$ system in the decay $X(3872)\ra\jpsi\pi^{+}\pi^{-}$~\cite{intro:1-9}. 
Due to the proximity of its mass to the $\bar{D}D^{*}$ production threshold, the $X(3872)$ has been interpreted as a hadronic molecule or a tetraquark state~\cite{intro:1-10}. Since the $X(3872)$ is a $1^{++}$ state, it could be produced through the radiative transition of an excited vector charmonium or charmonium-like state. The process $\ee\ra\gamma X(3872)\ra\gamma\pi^{+}\pi^{-}\jpsi$ was first observed with a statistical significance of $6.3\sigma$~\cite{BESIII:3872}, with data samples collected by the BESIII, 
at $\ee$ center-of-mass (CM) energies from $\sqrt{s}=$ 4.009~GeV to 4.420~GeV~\cite{intro:1-17}. 
Understanding the properties of higher-mass conventional charmonium states is crucial for advancing knowledge of quantum chromodynamics~\cite{lilun1, lilun2}. 

Under the molecular hypothesis of the $X(3872)$, the existence of an S-wave $\Dsz\Dszb$ bound state $X_{2} [J^{PC}=2^{++}]$ is predicted in the effective field theory approach~\cite{intro:2-7, intro:2-8}. As a result of the heavy quark spin symmetry, the expected binding energy of the $X_{2}$ resonance is found to be similar to that of the $X(3872)$~\cite{3872-4013}, $i.e.$,
\begin{linenomath*}
\begin{equation}
 M_{X_{2}} - M_{X(3872)} \approx M_{D^{*}} - M_{D} \approx 140~\mevcc. 
 \label{eq-3872-4013}
\end{equation}
\end{linenomath*}
Therefore, the $X_{2}$ is also called $X(4013)$. 
The existence of such a state is also suggested in Refs.~\cite{intro:2-9, intro:2-10, intro:2-11, intro:2-12, intro:2-13, intro:2-14}. 

Recently, the process $\gamma\gamma\ra\gamma\psip$ 
is studied in the $\gamma\gamma$ mass range from the threshold to $4.200~\gevcc$ by Belle, 
and a structure $R$ is seen
in the invariant mass distribution of $\gamma\psip$ with the mass of $M_{R} = 4014.3 \pm 4.0 \pm 1.5~\mevcc$ and the width of $\Gamma_{R} = 4 \pm 11 \pm 6$ MeV~\cite{par:x4013}. The mass of $R$ agrees with the heavy quark spin symmetry predicted mass ($\approx 4013~\mevcc$) of the $X(4013)$~\cite{intro:3-18}. 
Meanwhile, the width of $R$ coincides with the predicted width of $2-8$ MeV for the $X(4013)$~\cite{3872-4013}. Thus, it is thought to be the $X(4013)$. However the global significance of $R$ was determined to be $2.8\sigma$ indicating that additional studies on the X are necessary. Meanwhile, the mass threshold of $\Dsz\Dszb$ is close to the mass of $X(4013)$~\cite{pdg}. Therefore in this analysis, we search for the decay $\dx$ in the process $\ee\ra\gamma X$. 

In addition, according to the theoretical prediction in Ref.~\cite{par:other-four}, $\etac$, $\chiczp$, $\chicop$ and $\chictp$ should be observable in $\Dsz\Dszb$ final states. Therefore, in this paper, we also search for these states in the process $\ee\ra\gamma X$. 

The final states $\DszDszb$ are reconstructed in the decays $\Dsz\Dszb\ra\piz\piz\Dz\Dzb$, and $\Dsz\Dszb\ra\gamma\piz\Dz\Dzb$. The first process carries 0.42 of the total branching fraction and the second 0.46~\cite{pdg}. 
The $\Dsz\Dszb\ra\gamma\gamma\Dz\Dzb$ final state is not included due to the smaller branching fraction of $\approx$0.12~\cite{pdg}.

\section{BESIII DETECTOR, DATA SAMPLE, AND MONTE CARLO SIMULATION}

The BESIII detector~\cite{bes3:detector} records symmetric $e^+e^-$ collisions provided by 
the BEPCII storage ring~\cite{bes3:detector2} in the CM energy range from 1.84 to 4.95~GeV, with a peak luminosity of $1.1 \times 10^{33}\;\text{cm}^{-2}\text{s}^{-1}$ 
achieved at $\sqrt{s} = 3.773\;\text{GeV}$. BESIII has collected large data samples in this 
energy region~\cite{bes3:detector3,EcmsMea,EventFilter}. 
The cylindrical core of the BESIII detector covers 93\% of the full solid angle and consists of 
a helium-based multilayer drift chamber~(MDC), a plastic scintillator time-of-flight system~(TOF), 
and a CsI(Tl) electromagnetic calorimeter~(EMC), which are all enclosed in a superconducting 
solenoidal magnet providing a 1.0~T magnetic field. The solenoid is supported by an octagonal 
flux-return yoke with resistive plate counter muon identification modules interleaved with steel. 
The charged-particle momentum resolution at $1~{\rm GeV}/c$ is $0.5\%$, and the ${\rm d}E/{\rm d}x$ 
resolution is $6\%$ for electrons from Bhabha scattering. The EMC measures photon energies with 
a resolution of $2.5\%$ ($5\%$) at $1$~GeV in the barrel (end cap) region. The time resolution 
in the TOF barrel region is 68~ps, while that in the end cap region was 60~ps~\cite{bes3:detector4}.

This analysis use data samples corresponding to an integrated luminosity of 1667.4~$\rm pb^{-1}$, collected at a CM energy of $\sqrt{s} = 4.682$~GeV with the BESIII detector at the BEPCII. 
Simulated samples produced with a {\sc geant4}-based~\cite{bes3:g4} Monte Carlo (MC) package, which includes the geometric description of the BESIII detector and the detector response, are used to determine the detection efficiency and to estimate the backgrounds. 
The simulation models the beam energy spread and initial state radiation in the $e^+e^-$ annihilations with the generator {\sc kkmc}~\cite{bes3:kkmc}. 
The inclusive MC sample used in this study corresponds to ten times the integrated luminosity of data, generated at $\sqrt{s}=4.682$~GeV, which includes the production of open charm processes, the ISR production of vector charmonium(-like) states, and the continuum processes incorporated in {\sc kkmc}~\cite{bes3:kkmc}. 
All particle decays are modeled with {\sc evtgen}~\cite{bes3:evtgen} using branching fractions (BFs) either taken from the Particle Data Group~(PDG)~\cite{pdg}, when available, or otherwise estimated with {\sc lundcharm}~\cite{bes3:lundcharm}. Final state radiation from charged final state particles is incorporated using the {\sc photos} package~\cite{bes3:photon}.

In this analysis,  
the exclusive processes of $\ee\to\gamma X$, where  $X$ refers to $\etac$ or $\chicjp$ are generated with an angular distribution of $(1+\lambda \rm{cos}^{2}\theta)$, where $\theta$ is the polar angle between the radiative photon and the positron beam direction in the center-of mass system, and the value of $\lambda$ is set to 1 
for $\etac$ and to 1, -1/3, 1/13 for $\chiczp$, $\chicop$ and $\chictp$ respectively\cite{JPE, P2GC}. The exclusive processes of $\ee\to\gamma X$ are generated in phase space, where $X$ refers to $\x$. 
Considering the background level and the expected number of signal events, the decays $\konepi$, $\kthreepi$ and $\kspipi$ are studied. Charge conjugation is implied. These three decay modes of the $\Dz$ are combined into nine decay-chain categories : \\
DC-11: $\Dz\Dzb\ra (K^{-}\pi^{+})(K^{+}\pi^{-})$, \\
DC-12: $\Dz\Dzb\ra (K^{-}\pi^{+})(K^{+}\pi^{-}\pi^{-}\pi^{+})$, \\
DC-13: $\Dz\Dzb\ra (K^{-}\pi^{+})(K_{S}^{0}\pi^{+}\pi^{-})$, \\
DC-21: $\Dz\Dzb\ra (K^{-}\pi^{+}\pi^{+}\pi^{-})(K^{+}\pi^{-})$, \\
DC-22: $\Dz\Dzb\ra (K^{-}\pi^{+}\pi^{+}\pi^{-})(K^{+}\pi^{-}\pi^{-}\pi^{+})$, \\
DC-23: $\Dz\Dzb\ra (K^{-}\pi^{+}\pi^{+}\pi^{-})(K_{S}^{0}\pi^{+}\pi^{-})$, \\
DC-31: $\Dz\Dzb\ra (K_{S}^{0}\pi^{+}\pi^{-})(K^{+}\pi^{-})$, \\
DC-32: $\Dz\Dzb\ra (K_{S}^{0}\pi^{+}\pi^{-})(K^{+}\pi^{-}\pi^{-}\pi^{+})$, and \\
DC-33: $\Dz\Dzb\ra (K_{S}^{0}\pi^{+}\pi^{-})(K_{S}^{0}\pi^{+}\pi^{-})$. \\
Two groups of MC samples (I, II) are generated for the signal processes. Group-I is produced with the predicted masses and widths of the $X$ states~\cite{par:x4013, par:other-four}, as listed in Table~\ref{xmass}, while Group-II is with the masses ranging from 4.010 to
4.320 $\gevcc$ in steps of 0.01 $\gevcc$ and widths varying in the set of values: 0.005, 0.01, 0.03, 0.05, 0.07, and 0.09~GeV. The quantum numbers $J^{PC}$ of $X$ used in Group-I and Group-II are listed in Table~\ref{xmass}.

\begin{table}[htbp] 
\renewcommand{\arraystretch}{1.3}
	\caption{Parameters of the five $X$ states for the Group-I sample~\cite{par:x4013, par:other-four}. }\label{five-X}
	\begin{center}    
		\begin{tabular}{c|c|c|c} 
			\hline
			~~~$X$ meson~~ & ~~Mass ($\gevcc$)~~ & ~~Width (GeV)~~  & ~~$J^{PC}$~~~ \\\hline 
            $\x$      & 4.013  & 0.005  &  $2^{++}$  \\
            $\etac$   & 4.043  & 0.080  &  $0^{-+}$  \\
            $\chiczp$ & 4.202  & 0.051  &  $0^{++}$  \\
            $\chicop$ & 4.271  & 0.039  &  $1^{++}$  \\
            $\chictp$ & 4.317  & 0.066  &  $2^{++}$  \\\hline
            
		\end{tabular}
	\end{center}
	\label{xmass}
\end{table}

Additionally, to investigate the background level of this process, the simulated MC samples of the reactions $\ee\to\Dsz\Dszb$ and $\ee\to\Dz\Dszb\piz$ are generated with the line-shape of cross section quoted from Refs.~\cite{bkg1:1, bkg2:1}. 
In the MC simulation, the $\Dsz$ and $\Dz$ meson decay into the same final states as the signal MC samples.

\section{EVENT SELECTION}
The candidate events for $\ee\ra\gamma X\ra\gamma\Dsz\Dszb$ are selected with a full reconstruction method. Charged tracks detected in the MDC are required to be within a polar angle range of $|\cos\theta|<0.93$, where the angle $\theta$ is defined with respect to the $z$-axis, which is the symmetry axis of the MDC. 
For charged tracks not originating from $K_{S}^{0}$ decays, the distance along the $z$-axis $|V_{z}|$ of the closest approach to the interaction point must be less than 10 cm, and that in the transverse plane $|V_{xy}|$ less than 1~cm. Particle identification (PID) for charged tracks combines measurements of the energy deposited in the MDC and the flight time in the TOF to form likelihoods $\mathcal{L}(h)~(h=K, \pi)$ for each hadron $h$ hypothesis. Tracks are identified as kaons (pions) when the kaon (pion) hypothesis has larger probability $\mathcal{L}(K) > \mathcal{L}(\pi)$ ($\mathcal{L}(\pi) > \mathcal{L}(K)$). Tracks without valid PID information are rejected. 

Each $K_{S}^{0}$ candidate is reconstructed from two oppositely charged tracks satisfying $|V_{z}| \leq 20$ cm and $|\rm{cos}\theta| \leq 0.93$. The two charged tracks are assumed to be a $\pp$ pair without imposing PID criteria. They are constrained to originate from a common vertex and are required to have an invariant mass $M_{\pp}$ such that: $0.45~\gevcc \leq M_{\pp} \leq 0.55~\gevcc$. 
The $\chi^{2}$ of the vertex fit is required to be less than 100. A secondary vertex fit is performed to ensure that the $K_{S}^{0}$ momentum points back to the interaction point. The decay length of the $K_{S}^{0}$ candidate is required to be greater than twice of the vertex resolution.

Photon candidates are identified using showers in the EMC. The deposited energy of each shower must be more than $0.025~\gev$ in the barrel ($|\cos\theta|<0.8$) and more than $0.050~\gev$ in the end cap ($0.86<|\cos\theta|<0.92$) regions. To suppress electronic noise and energy depositions unrelated to the event, the difference between the EMC cluster time and the reconstructed event start time is required to be within $[0,~700]$~ns. 
To veto showers originating from charged tracks, the opening angle between the extrapolated trajectory of any charged track and the shower position must exceed 10 degrees. 

The selected $K^{\pm}$, $\pi^{\pm}$, and $K_{S}^{0}$ candidates in the event are combined to reconstruct $\Dz\ra K \pi$, $\Dz\ra K \pi^{+}\pi^{-}\pi$ and $\Dz\ra K_{S}^{0}\pi^{+}\pi^{-}$ decays. The $\Dz$ candidates are kept if the invariant masses of the $K \pi$, $K \pi^{+}\pi^{-}\pi$ or $K_{S}^{0}\pi^{+}\pi^{-}$ systems, denoted as $M_{\Dz}$, are within a 150 $\mevcc$ mass window around the nominal $\Dz$ mass. 
Each $K^{\pm}$, $\pi^{\pm}$, and $K_{S}^{0}$ candidate is used at most once in an event. If there are multi-combinations in an event, the one with $(M_{\Dz}+M_{\Dzb})/2$ closest to the nominal mass of the $\Dz$ meson~\cite{pdg} is chosen.

In addition to the $\Dz\Dzb$ pair, for the process $\ee\ra\gamma\Dsz\Dszb\ra\gamma\piz\piz\Dz\Dzb$, an event is required to contain at least 5 good photon candidates. By looping over all the photon candidates, a 1-constraint (1C) kinematic fit is performed for two photons and only combinations satisfying $0.1 < m_{\gamma\gamma} < 0.17~\gevcc$, $P < 0.14$~GeV/$c$ and $\chi^{2}_{\rm 1C} < 20$ are regarded as $\piz$ candidates, in which $m_{\gamma\gamma}$ is the unconstrained invariant mass of $\gamma\gamma$, $P$ is the modulus of the momentum of the $\gamma\gamma$ pair, and the $\chi^{2}_{\rm 1C}$ is given by the 1C kinematic fit. 
After obtaining the $\piz$ candidates, a 4-constraint (4C) kinematic fit is performed for all the final states. 
If more than two $\piz$ candidates and (or) more than one $\Dz\Dzb$ pair decaying in different decay channels are found in an event, the combination with the minimum $\chi^{2}_{\rm tot}$ is retained. The $\chi^{2}_{\rm tot}$ is calculated from 4C kinematic fit and 1C kinematic fit of $\pi^{0}_{1}$ and $\pi^{0}_{2}$ $\left( \chi^{2}_{\rm tot} = \chi^{2}_{\rm 4C} + \chi^{2}_{\rm 1C}(\pi^{0}_{1}) + \chi^{2}_{\rm 1C}(\pi^{0}_{2}) \right)$, in which $\chi^{2}_{\rm 1C}(\pi^{0}_{1})$ and $\chi^{2}_{\rm 1C}(\pi^{0}_{2})$ are given by the 1C kinematic fit of the first and second pion candidate $\pi^{0}_{1}$ and $\pi^{0}_{2}$. With minimum $\chi^{2}_{\rm tot}$, the final state particles are also determined. The selected $K^{\pm}$, $\pi^{\pm}$ and $\ks$ candidates in the event are combined to reconstruct the $\Dz$ meson through its decays $\konepi$, $\kthreepi$ and $\kspipi$. Each one of the two selected $\piz$ mesons is combined with the $\Dz$ or $\Dzb$. The combinations with the mass closest to the nominal mass of the $\Dsz$ meson~\cite{pdg} are identified as $\Dsz$ and $\Dszb$.  

For the $\ee\ra\gamma\Dsz\Dszb\ra\gamma\gamma\piz\Dz\Dzb$ process, each event is required to contain at least 4 good photon candidates, and the same 1C kinematic fit is performed for the photon pairs to determined $\piz$ candidates. The best photon candidates are selected with minimum $\chi^{2}_{\rm tot}$ calculated from 4C kinematic fit and 1C kinematic fit for $\piz$ $\left( \chi^{2}_{\rm tot} = \chi^{2}_{\rm 4C} + \chi^{2}_{\rm 1C}(\piz) \right)$. With minimum $\chi^{2}_{\rm tot}$, the final state particles are also determined. The selected $K^{\pm}$, $\pi^{\pm}$ and $\ks$ candidates in the event are combined to reconstruct the $\Dz$ meson through its decays $\konepi$, $\kthreepi$ and $\kspipi$. The selected $\piz$ meson and each one of the two final photons are combined with the $\Dz$ or $\Dzb$, the combinations with the mass closest to the nominal mass of the $\Dsz$ meson~\cite{pdg} are selected as $\Dsz$ and $\Dszb$. 

To suppress backgrounds, only the events in the $\Dz$, $\Dzb$, $\Dsz$ and $\Dszb$ mass windows are kept for further analyses. The mass window regions are listed in Table~\ref{masswindow}, and correspond to three times their mass resolutions. 

\begin{table}[htbp] 
\renewcommand{\arraystretch}{1.3}
	\caption{The mass windows for $\Dz$, $\Dzb$, $\Dsz$ and $\Dszb$ candidates.}
	\begin{center}    
		\begin{tabular}{ c  c  c }  
			\hline
\multicolumn{3}{c}{$\ppDD$ mode}\\\hline
~~~ ~~~  ~&~ ~~~~$M_{\Dz}$ ($\gevcc$) ~~~&~~~ $M_{\Dzb}$ ($\gevcc$)~~~ \\\hline
$\x$          ~&~ (1.845, 1.887) ~&~ (1.844, 1.887) \\
$\etac$       ~&~ (1.842, 1.890) ~&~ (1.846, 1.886)  \\
$\chiczp$     ~&~ (1.846, 1.885) ~&~ (1.844, 1.888)  \\
$\chicop$     ~&~ (1.847, 1.886) ~&~ (1.843, 1.890)  \\
$\chictp$     ~&~ (1.847, 1.886) ~&~ (1.844, 1.889) 
\\\hline
~    ~&~ $M_{\Dsz}$ ($\gevcc$) ~&~ $M_{\Dszb}$ ($\gevcc$) \\\hline
$\x$          ~&~ (1.984, 2.034) ~&~ (1.983, 2.033) \\
$\etac$       ~&~ (1.983, 2.035) ~&~ (1.982, 2.035)  \\
$\chiczp$     ~&~ (1.983, 2.037) ~&~ (1.982, 2.036)  \\
$\chicop$     ~&~ (1.983, 2.037) ~&~ (1.982, 2.036)  \\
$\chictp$     ~&~ (1.985, 2.034) ~&~ (1.981, 2.037) 
\\\hline
\multicolumn{3}{c}{$\gampiDD$ mode}\\\hline
~    ~&~ $M_{\Dz}$ ($\gevcc$) ~&~ $M_{\Dzb}$ ($\gevcc$) \\\hline
$\x$          ~&~ (1.849, 1.885) ~&~ (1.846, 1.888) \\
$\etac$       ~&~ (1.848, 1.886) ~&~ (1.848, 1.887)  \\
$\chiczp$     ~&~ (1.847, 1.887) ~&~ (1.850, 1.885)  \\
$\chicop$     ~&~ (1.846, 1.889) ~&~ (1.847, 1.888)  \\
$\chictp$     ~&~ (1.848, 1.887) ~&~ (1.847, 1.888) 
\\\hline
~    ~&~ $M_{\Dsz}$ ($\gevcc$) ~&~ $M_{\Dszb}$ ($\gevcc$) \\\hline
$\x$          ~&~ (1.989, 2.025) ~&~ (1.989, 2.025) \\
$\etac$       ~&~ (1.989, 2.027) ~&~ (1.988, 2.029)  \\
$\chiczp$     ~&~ (1.988, 2.029) ~&~ (1.983, 2.035)  \\
$\chicop$     ~&~ (1.988, 2.029) ~&~ (1.984, 2.032)  \\
$\chictp$     ~&~ (1.989, 2.028) ~&~ (1.982, 2.036) 
\\\hline

		\end{tabular}
	\end{center}
	\label{masswindow}
\end{table}

\section{BACKGROUND ESTIMATION}
The background contributions are studied using both inclusive and exclusive MC samples, where exclusive MC samples are used to model specific background components. They are decomposed into four components: $\ee\ra\Dsz\Dszb$ background ($B_{1}$), $\ee\ra\Dz\Dszb\piz$ background ($B_{2}$), cross-feed background from one signal channel to the other ($B_{3}$) and other combinatorial backgrounds ($B_{4}$).

\subsection{$\ee\ra\Dsz\Dszb$ and $\ee\ra\Dz\Dszb\piz$ background}
The number of background events from $\ee\ra\Dsz\Dszb$ ($N_{B_{1}}$) and the number of background events from $\ee\ra\Dz\Dszb\piz$ ($N_{B_{2}}$) used in the final fits are calculated as 
\begin{linenomath*}
\begin{equation}
	N_{B_{1}} = \mathscr{L} \sigma_{1} \bar{\epsilon_{b1}} \sum_{m,n=1}^{2} \sum_{i,j=1}^{3} {\rm Br}_{m} {\rm Br}_{n} {\rm Br}_{i} {\rm Br}_{j},
	\label{eq-bkg-num1}
\end{equation}
\end{linenomath*}
\begin{linenomath*}
\begin{equation}
	N_{B_{2}} = \mathscr{L} \sigma_{2} \bar{\epsilon_{b2}} \sum_{m=1}^{2} \sum_{i,j=1}^{3} {\rm Br}_{m} {\rm Br}_{i} {\rm Br}_{j} ,
	\label{eq-bkg-num2}
\end{equation}
\end{linenomath*}
where $\mathscr{L}$ is the integrated luminosity at $\sqrt{s}=4.682$~GeV~\cite{luminosity_4680}, $\sigma_{1}$ and $\sigma_{2}$ are the cross sections of the $\ee\ra\Dsz\Dszb$ and $\ee\ra\Dz\Dszb\piz$\cite{bkg1:1, bkg2:1} processes, 
${\rm Br}_{m/n=1}$ is the products of the BFs of $\dpizDz$ and $\piz\ra\gamma\gamma$, ${\rm Br}_{m/n=2}$ is the branching fraction of the radiative decay $\dgamDz$~\cite{pdg}, ${\rm Br}_{i/j=1,2}$ are the BFs of $\Dz\ra\kmpip$ and $\Dz\ra\kmpipipi$, ${\rm Br}_{i/j=3}$ is the product of the BFs of $\Dz\ra\kszpipi$ and $K_{S}^{0}\ra\pi^{+}\pi^{-}$~\cite{pdg}, and $\bar{\epsilon_{b1}}$ and $\bar{\epsilon_{b2}}$ are the average detection efficiencies of $\ee\ra\Dsz\Dszb$ and $\ee\ra\Dz\Dsz\piz$ after all selections, respectively. The uncertainties of $N_{B_{1}}$ and $N_{B_{2}}$ are calculated by varying the corresponding values by $\pm 1\sigma$. The corresponding detection efficiencies and the numbers of background events are summarized in Table~\ref{bkg-num}.

\subsection{Cross-feed background from the one signal channel to the other}
The cross-feed background $N_{B_{3}}$ indicates the mutual influence between two signal processes $\ee\ra\gamma\Dsz\Dszb\ra\gamma\piz\piz\Dz\Dzb$ and $\ee\ra\gamma\Dsz\Dszb\ra\gamma\gamma\piz\Dz\Dzb$. 
The process $\ee\ra\gamma\Dsz\Dszb\ra\gamma\gamma\piz\Dz\Dzb$ 
is the cross-feed background for the signal process $\ee\ra\gamma\Dsz\Dszb\ra\gamma\piz\piz\Dz\Dzb$ and vice versa. 
According to the exclusive MC sample, for the $\ppDD$ mode, the selection efficiencies for the cross-feed background $\ee\ra\gampiDD$ are 0.31\%, 0.35\%, 0.27\%, 0.23\% and 0.19\% for $\x$, $\etac$, $\chiczp$, $\chicop$ and $\chictp$, respectively. For the $\gampiDD$ mode, the selection efficiencies for the cross-feed background $\ee\ra\ppDD$ are 0.07\%, 0.13\%, 0.34\%, 0.34\% and 0.24\% for $\x$, $\etac$, $\chiczp$, $\chicop$ and $\chictp$, respectively. The numbers of these background events are constrained by the cross section obtained from this paper, the branching fraction, the integrated luminosity, and the corresponding efficiency. 

\subsection{Other smooth backgrounds}
The remaining background events $N_{B_{4}}$ are about 3.4\% for the 
$\ppDD$ mode and 5.2\% for the $\gampiDD$ mode, estimated using the inclusive MC sample. 
The line-shapes of these backgrounds are from the inclusive MC sample.

\begin{figure*}[!tbp]
\centering
\begin{overpic}[width=0.48\textwidth, height=5cm]{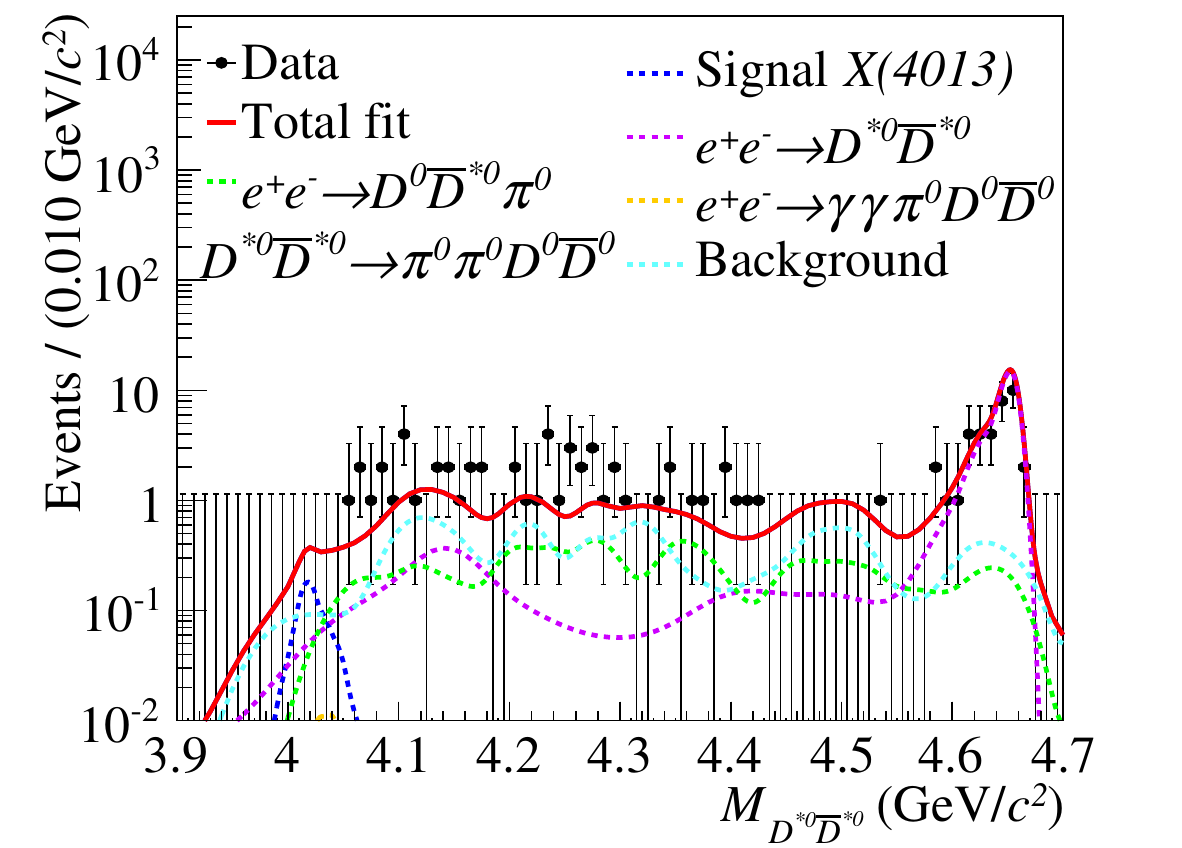}
  \put(50,1.2){(a)}
\end{overpic}\hfill
\begin{overpic}[width=0.48\textwidth, height=5cm]{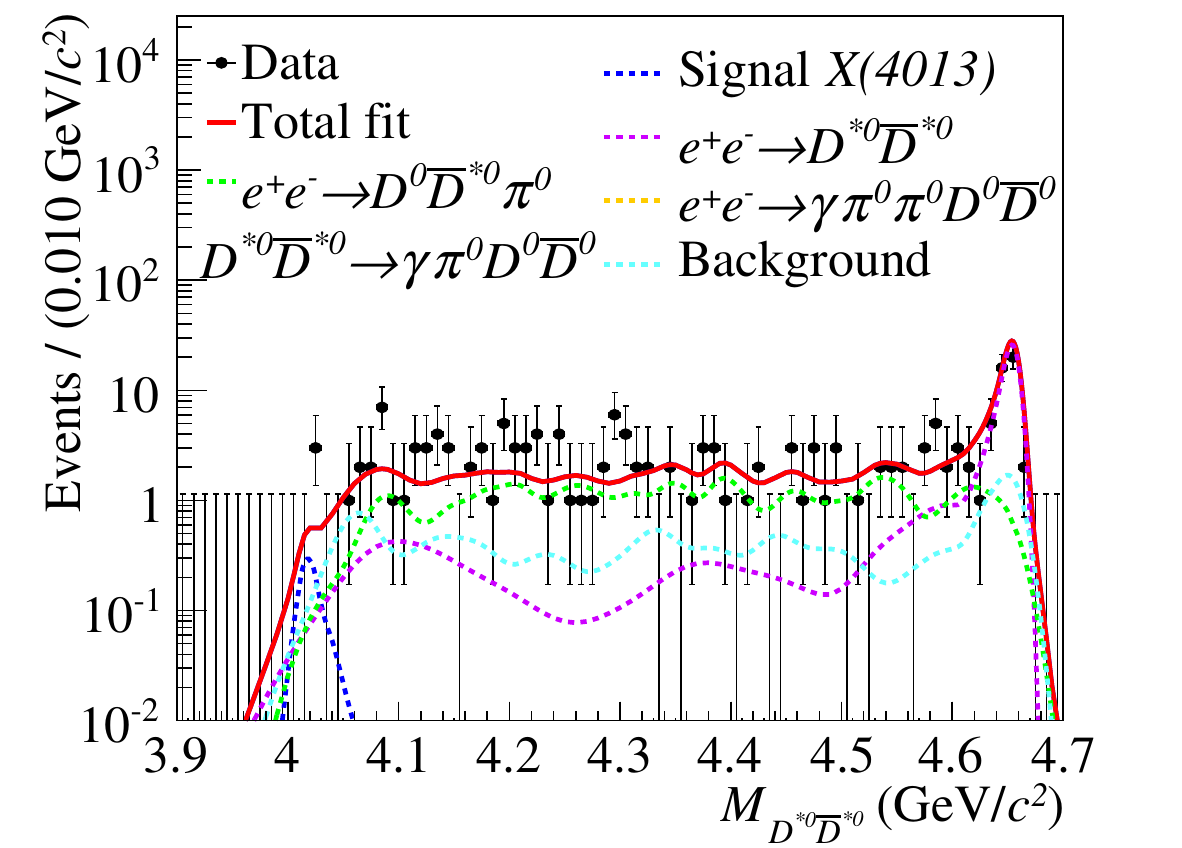}
  \put(50,1.2){(f)}
\end{overpic}
\par\vspace{0.10cm}

\begin{overpic}[width=0.48\textwidth, height=5cm]{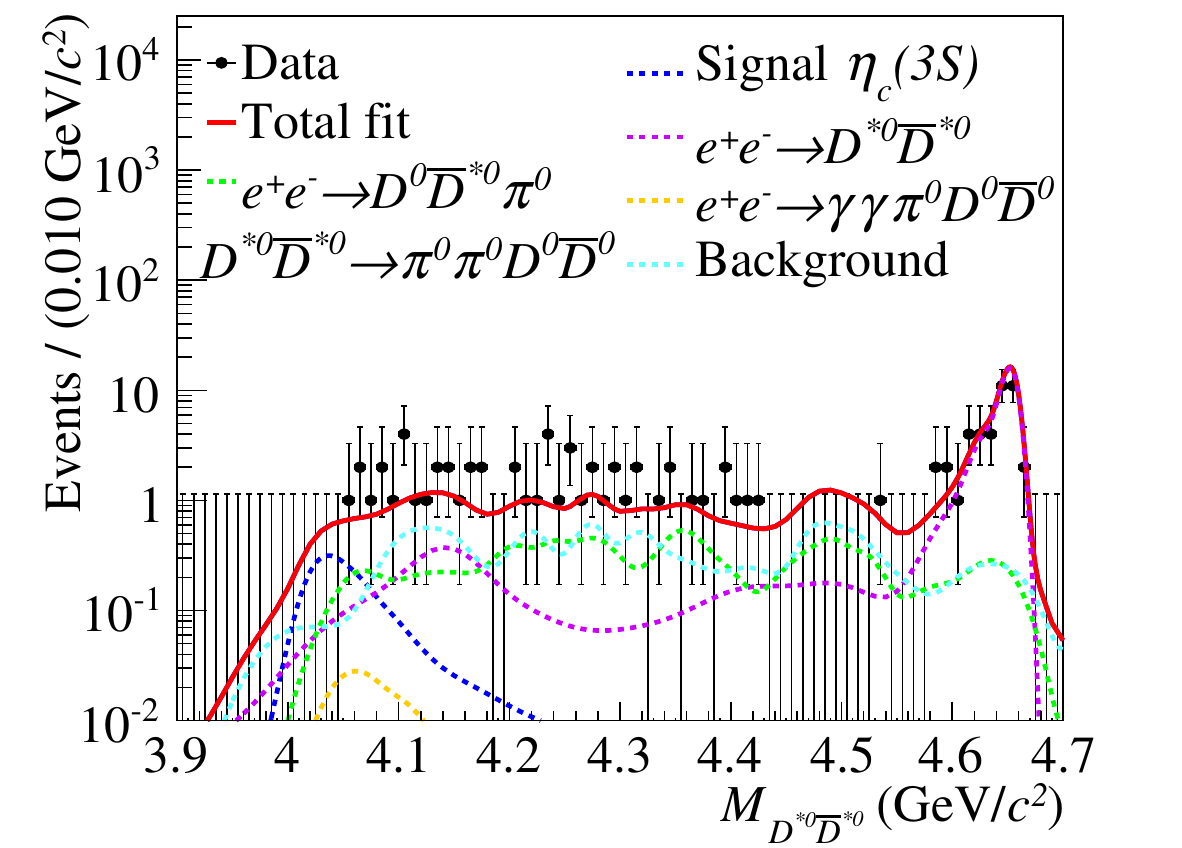}
  \put(50,1.2){(b)}
\end{overpic}\hfill
\begin{overpic}[width=0.48\textwidth, height=5cm]{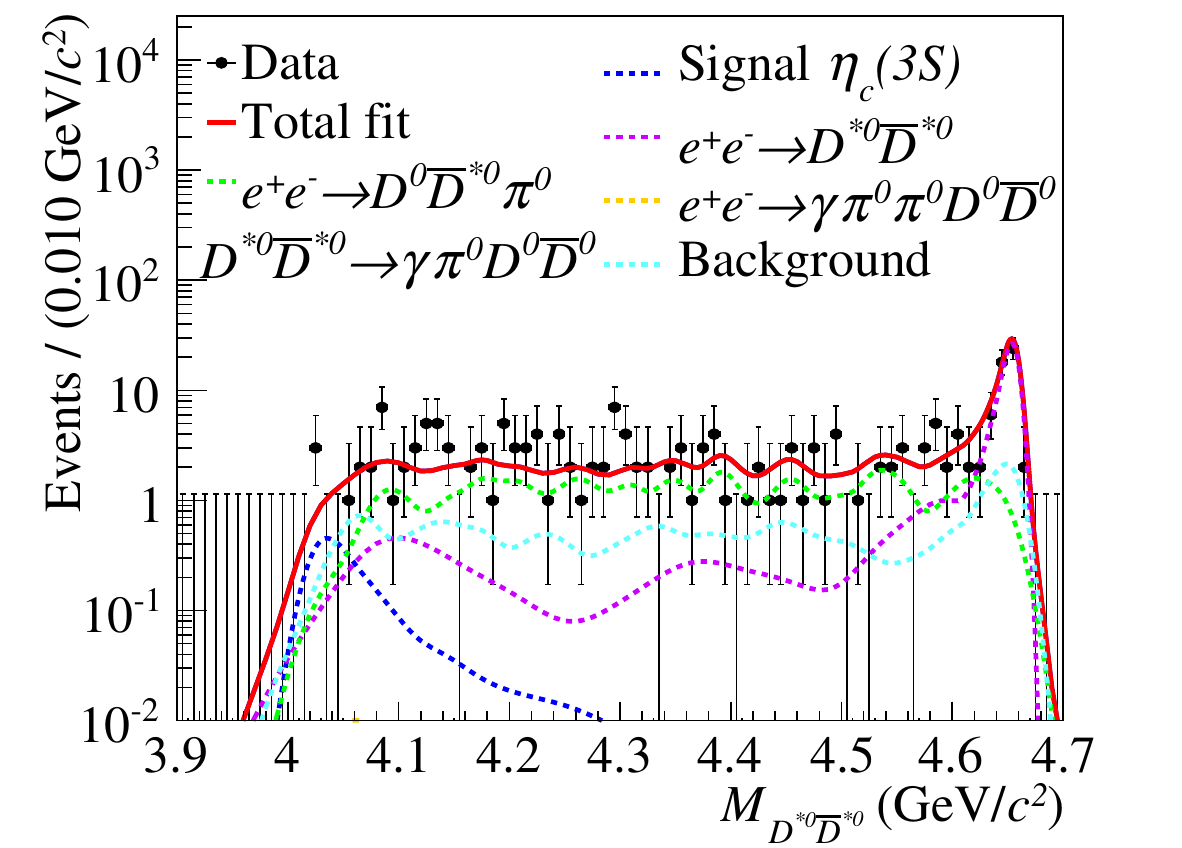}
  \put(50,1.2){(g)}
\end{overpic}
\par\vspace{0.10cm}

\begin{overpic}[width=0.48\textwidth, height=5cm]{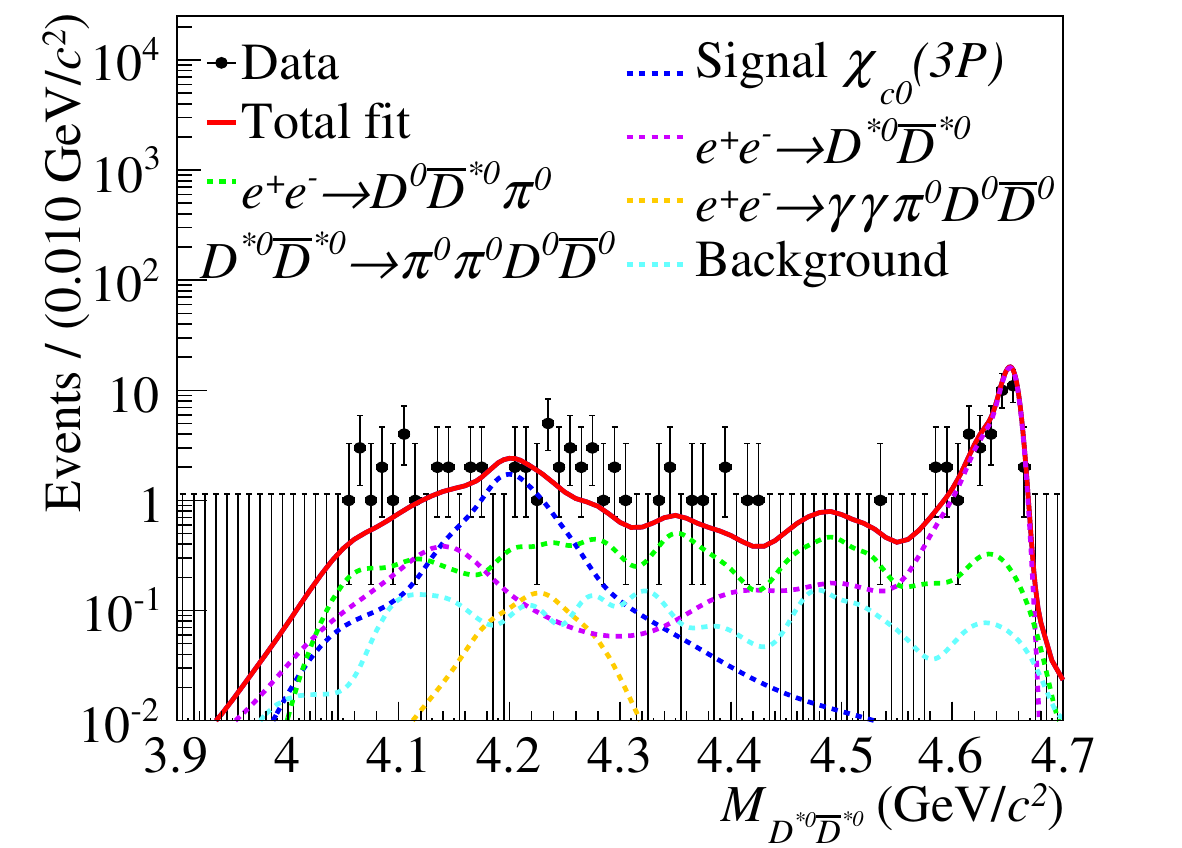}
  \put(50,1.2){(c)}
\end{overpic}\hfill
\begin{overpic}[width=0.48\textwidth, height=5cm]{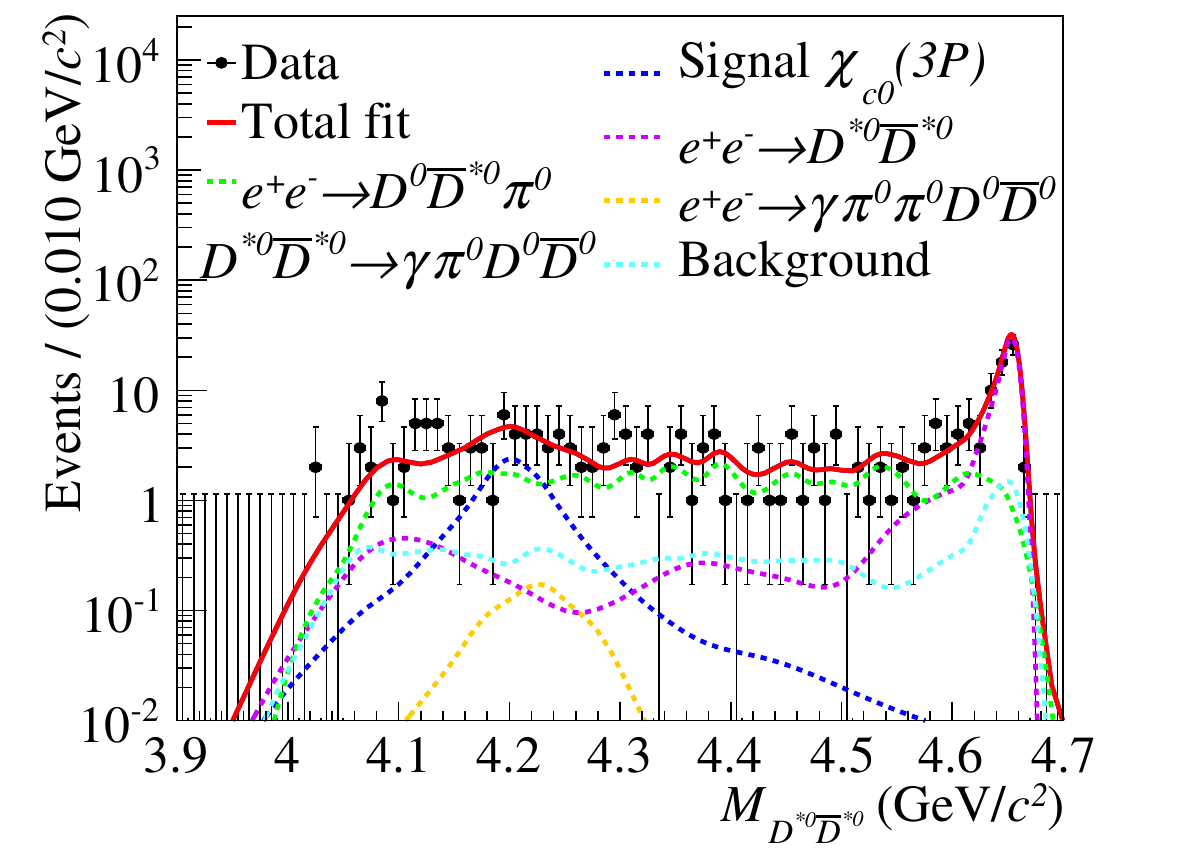}
  \put(50,1.2){(h)}
\end{overpic}
\par\vspace{0.10cm}

\begin{overpic}[width=0.48\textwidth, height=5cm]{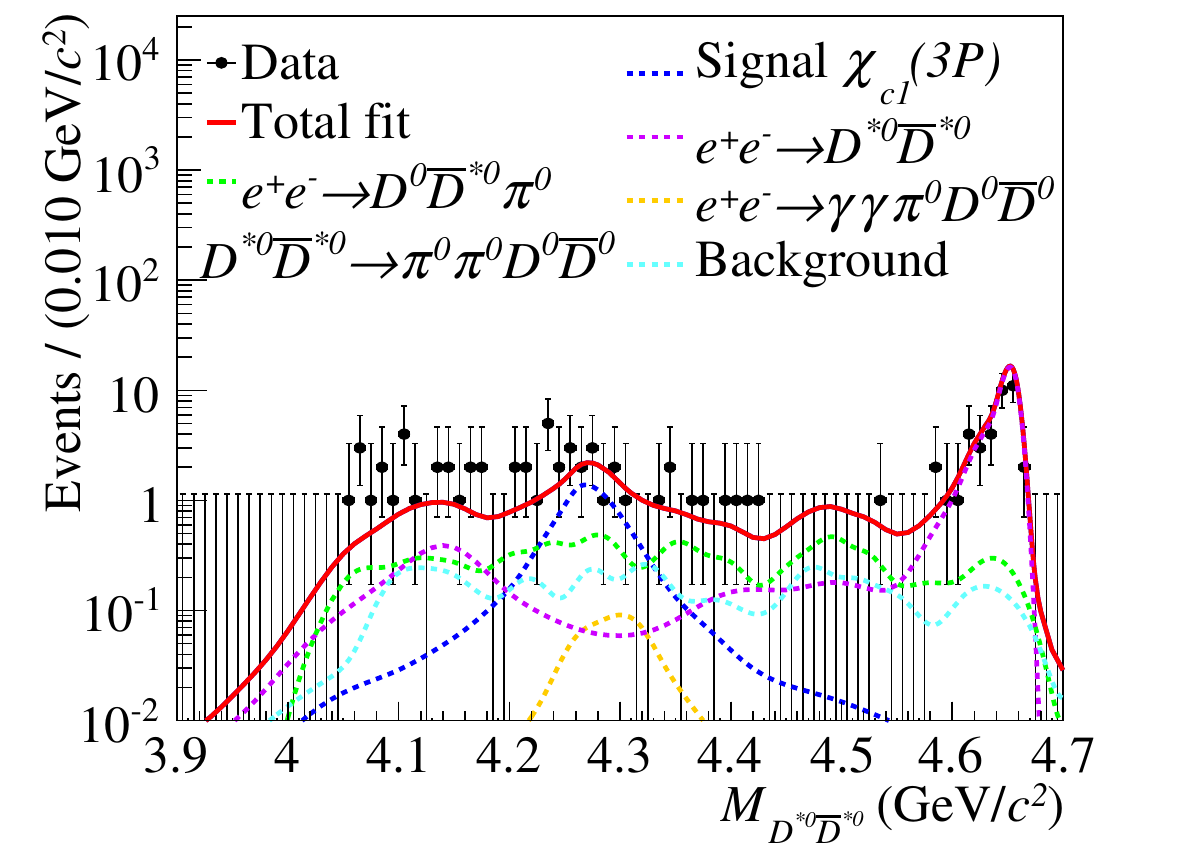}
  \put(50,1.2){(d)}
\end{overpic}\hfill
\begin{overpic}[width=0.48\textwidth, height=5cm]{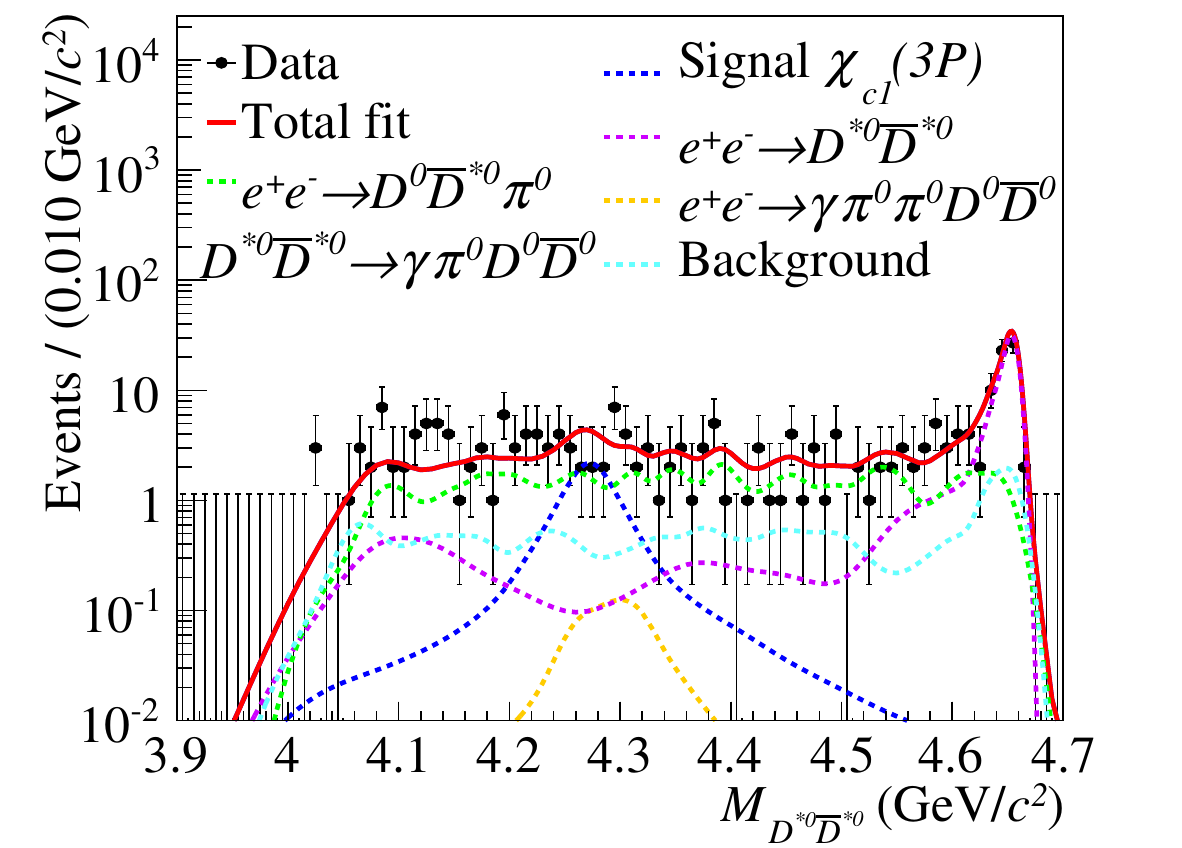}
  \put(50,1.2){(i)}
\end{overpic}

\begin{overpic}[width=0.48\textwidth, height=5cm]{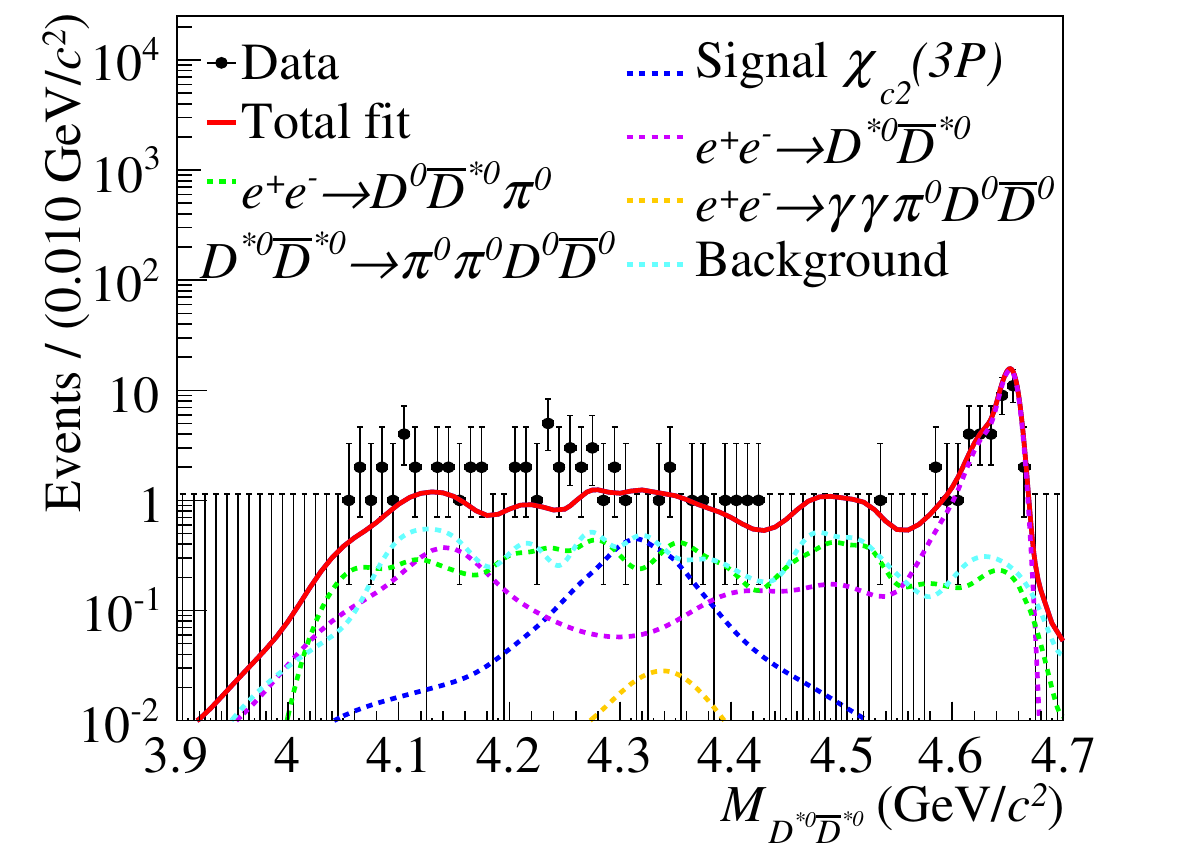}
  \put(50,1.2){(e)}
\end{overpic}\hfill
\begin{overpic}[width=0.48\textwidth, height=5cm]{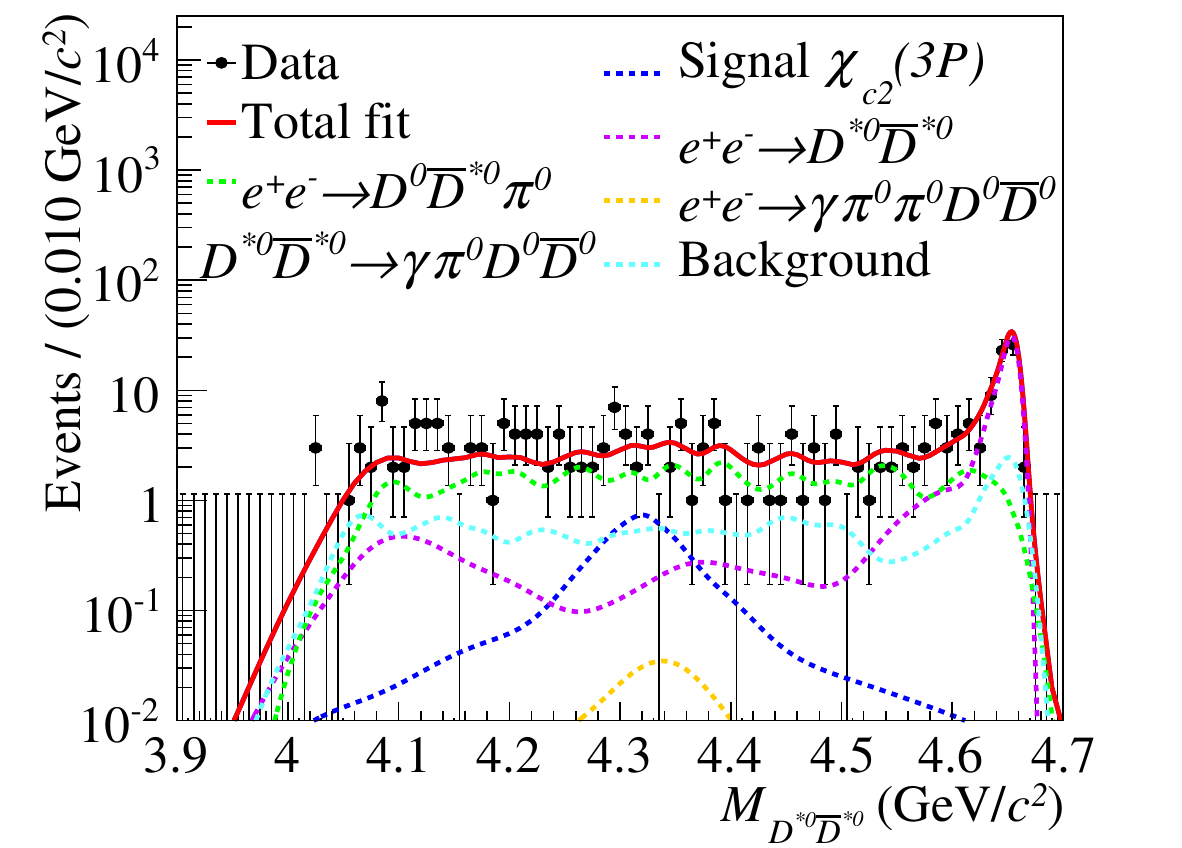}
  \put(50,1.2){(j)}
\end{overpic}
\end{figure*}

\begin{figure*}[!t]
\centering
\caption{The results of the fit to the $\Dsz\Dszb$ invariant mass distributions with the masses and widths set to the values given in Table~\ref{five-X}. The dots with error bars are the data distributions and the solid red lines are the best fit results. The dashed blue lines show the signal shapes, the dashed orange lines represent the contributions from cross-feed backgrounds, the dashed purple lines show the contributions from the $\ee\ra\Dsz\Dszb$, the dashed green lines are the contributions from the $\ee\ra\Dz\Dszb\piz$, and the dashed cyan lines represent the other backgrounds estimated using the inclusive MC sample. (a, b, c, d, e) for the $\ppDD$ mode and (f, g, h, i, j) for the $\gampiDD$ mode. Since the results of $(\sigma\cdot {\rm Br})$ obtained through fitting are relatively small for both $\x$ and $\etac$ (as listed in Table~\ref{sim}), the contributions from the cross-feed backgrounds in the dashed orange lines depend on the fitted cross section, and thus are also small.}

\label{simfit}
\end{figure*}

\FloatBarrier


\begin{table*}[htbp] 
\renewcommand{\arraystretch}{1.3}
	\caption{Summary of the detection efficiencies after all selections, $\bar{\epsilon_{b1}}$ and $\bar{\epsilon_{b2}}$, along with the corresponding numbers of background events, $N_{B_{1}}$ and $N_{B_{2}}$, for the $\ppDD$ and $\gampiDD$ modes, respectively. The former ($\bar{\epsilon_{b1}}$ and $N_{B_{1}}$) are derived from the background process $\ee\ra\Dsz\Dszb$, while the latter ($\bar{\epsilon_{b2}}$ and $N_{B_{2}}$) come from $\ee\ra\Dz\Dszb\piz$. 
    The uncertainties are calculated by varying the corresponding values used in Eq.~\ref{eq-bkg-num1} and Eq.~\ref{eq-bkg-num2} by $\pm 1\sigma$. }
	\begin{center}    
		\begin{tabular}{ c | c  c | c  c | c  c | c  c }  
			\hline
\multirow{2}*{~~~$X$ meson~~~}
			              & \multicolumn{4}{c|}{~~~~~~~~~~~~~$\ppDD$ mode ~~~~~~~~~~~~~}    &  \multicolumn{4}{c}{~~~~~~~~~~~~~$\gampiDD$ mode ~~~~~~~~~~~~~}    \\\cline{2-9}
                          & $\bar{\epsilon_{b1}}$ & $N_{B_{1}}$ & $\bar{\epsilon_{b2}}$ & $N_{B_{2}}$  & $\bar{\epsilon_{b1}}$ & $N_{B_{1}}$ & $\bar{\epsilon_{b2}}$ & $N_{B_{2}}$ \\\hline
$\x   $   ~&~ 0.44\%  ~&~ $    55.1 \pm 5.5$ ~&~ 0.13\%  ~&~ $    17.9 \pm 1.8$ ~&~ 0.56\% ~&~ $    71.9 \pm 5.7$  ~&~  0.50\% ~&~ $    71.5 \pm 5.7$\\
$\etac$   ~&~ 0.46\%  ~&~ $    57.7 \pm 5.7$ ~&~ 0.14\%  ~&~ $    20.4 \pm 2.0$ ~&~ 0.58\% ~&~ $    74.2 \pm 5.9$  ~&~  0.56\% ~&~ $    81.0 \pm 6.5$\\
$\chiczp$ ~&~ 0.46\%  ~&~ $    57.7 \pm 5.7$ ~&~ 0.15\%  ~&~ $    20.8 \pm 2.1$ ~&~ 0.68\% ~&~ $    86.9 \pm 6.9$  ~&~  0.67\% ~&~ $    97.0 \pm 7.8$\\
$\chicop$ ~&~ 0.46\%  ~&~ $    57.7 \pm 5.7$ ~&~ 0.14\%  ~&~ $    20.4 \pm 2.0$ ~&~ 0.70\% ~&~ $    89.2 \pm 7.1$  ~&~  0.66\% ~&~ $    95.4 \pm 7.6$\\
$\chictp$ ~&~ 0.44\%  ~&~ $    56.0 \pm 5.5$ ~&~ 0.14\%  ~&~ $    19.6 \pm 2.0$ ~&~ 0.69\% ~&~ $    88.1 \pm 7.0$  ~&~  0.69\% ~&~ $    99.0 \pm 7.9$\\\hline

		\end{tabular}
	\end{center}
	\label{bkg-num}
\end{table*}

\begin{table*}[htbp] 
\renewcommand{\arraystretch}{1.3}
	\caption{Results of the simultaneous fit and upper limits of $\sigma\cdot {\rm Br}$ for the five different signal modes. The uncertainties of $N^{\rm{sig}}$ and $(\sigma\cdot {\rm Br})$ are statistical. “$(\sigma\cdot {\rm Br})^{\rm{UL}}$ with sys.” stands for the upper limits at the 90\% confidence level with systematic uncertainties. } 
	\begin{center}    
		\begin{tabular}{ c | c | c |c | c | c   }  
			\hline
~~Channel  ~&~ $N^{\rm{sig}}$ ~&~ $(\sigma\cdot {\rm Br})$ (pb)~&~
            $\chi^{2}/N_{\rm{ndf}}$ ~&~ $(\sigma\cdot {\rm Br})^{\rm{UL}}$ (pb)  ~&~ $(\sigma\cdot {\rm Br})^{\rm{UL}}$ with sys. (pb)~ \\\hline
$\dx$      ~&~ $1.4^{+2.3}_{-1.6}$  ~&~ $1.2^{+1.9}_{-1.3}$      ~&~  1.9          ~&~  4   ~&~ 5.4 \\
$\detac$   ~&~ $6.0^{+5.8}_{-4.6}$  ~&~ $5.6^{+5.4}_{-4.4}$      ~&~  2.1          ~&~ 14   ~&~ 17.2\\
$\dchiczp$ ~&~ $39.7\pm11.1$ ~&~ $41.0\pm11.5$    ~&~  1.6          ~&~ 57  ~&~ 65.9 \\
$\dchicop$ ~&~ $28.4\pm9.4$ ~&~ $27.4\pm9.1$     ~&~  2.0          ~&~  40  ~&~ 44.5 \\
$\dchictp$ ~&~ $13.7\pm10.9$ ~&~ $14.2\pm11.3$    ~&~  2.1          ~&~  31  ~&~ 35.4 \\\hline

		\end{tabular}
	\end{center}
	\label{sim}
\end{table*}

\section{SIGNAL EXTRACTION}
The signal yields are determined from an unbinned extended maximum likelihood fit to the 
$M_{\Dsz\Dszb}$ spectra of the $\ppDD$ and $\gampiDD$ modes simultaneously. The probability density function (PDF) used in the fit is defined as 
\begin{linenomath*}
\begin{equation}
\mathrm{PDF}_{\rm total} = N^{\rm sig} S + \sum_{i=1}^{4} N_{B_{i}} B_{i}, 
\end{equation} 
\end{linenomath*}
where $i$ runs from 1 to 4 for the four background components, $N^{\rm sig}$ and $N_{B_{i}}$ represent the numbers of signal events and background events, and $S$ and $B_{i}$ are the signal and background PDFs. In the nominal fit, the signal and background PDF shapes are modeled with RooKeysPdf~\cite{keyspdf}. 
A smoothing parameter that controls the smoothing level of the RooKeysPdf is assigned a value of 0.4 for the $B_{1,2,4}$ to balance statistical fluctuations and over-smoothing. 

The signal PDF is taken from the signal MC. The background contributions are decomposed into the four components defined in section IV. The $B_{1,2,3}$ shapes 
are taken from the corresponding MC simulation, and the $B_{4}$ shape is taken from the inclusive MC sample. The parameter $N_{B_{4}}$ is free in the fit, $N_{B_{3}}$ depends on the product cross section obtained from this paper, while $N_{B_{1}}$ and $N_{B_{2}}$ are fixed.

In the simultaneous fit, $\sigma_{\ee\ra\gamma X}\cdot {\rm Br}_{X\ra\Dsz\Dszb}$ is calculated as 
\begin{linenomath*}
\begin{equation}
\begin{split}
	&\sigma\cdot {\rm Br} = \\
    & \dfrac{N^{\rm sig}}{\mathscr{L}} 
    \left[ ({\rm Br}_{n=1}\bar{\epsilon}_{1} + 
    2 {\rm Br}_{n=2} \bar{\epsilon}_{2}) {\rm Br}_{m=1} \sum_{i,j=1}^{3}{\rm Br}_{i} {\rm Br}_{j} \right]^{-1}, 
	\label{eqkedr}
    \end{split}
\end{equation}
\end{linenomath*}
where the symbols are those of Eq.~\ref{eq-bkg-num1} and Eq.~\ref{eq-bkg-num2}, while $\sigma\cdot {\rm Br}$ is the product of the cross section $\sigma_{\ee\ra\gamma X}$ and branching fraction ${\rm Br}_{X\ra\Dsz\Dszb}$, $\bar{\epsilon}_{1}$ and $\bar{\epsilon}_{2}$ are the average detection efficiencies for different $\Dz\Dzb$ decay modes in the $\ppDD$ and $\gampiDD$ modes, respectively. 
In the calculation, the BFs are taken from the world average value~\cite{pdg}.

\begin{figure*}[htbp] 
\centering
  \begin{overpic}[width=0.32\textwidth]{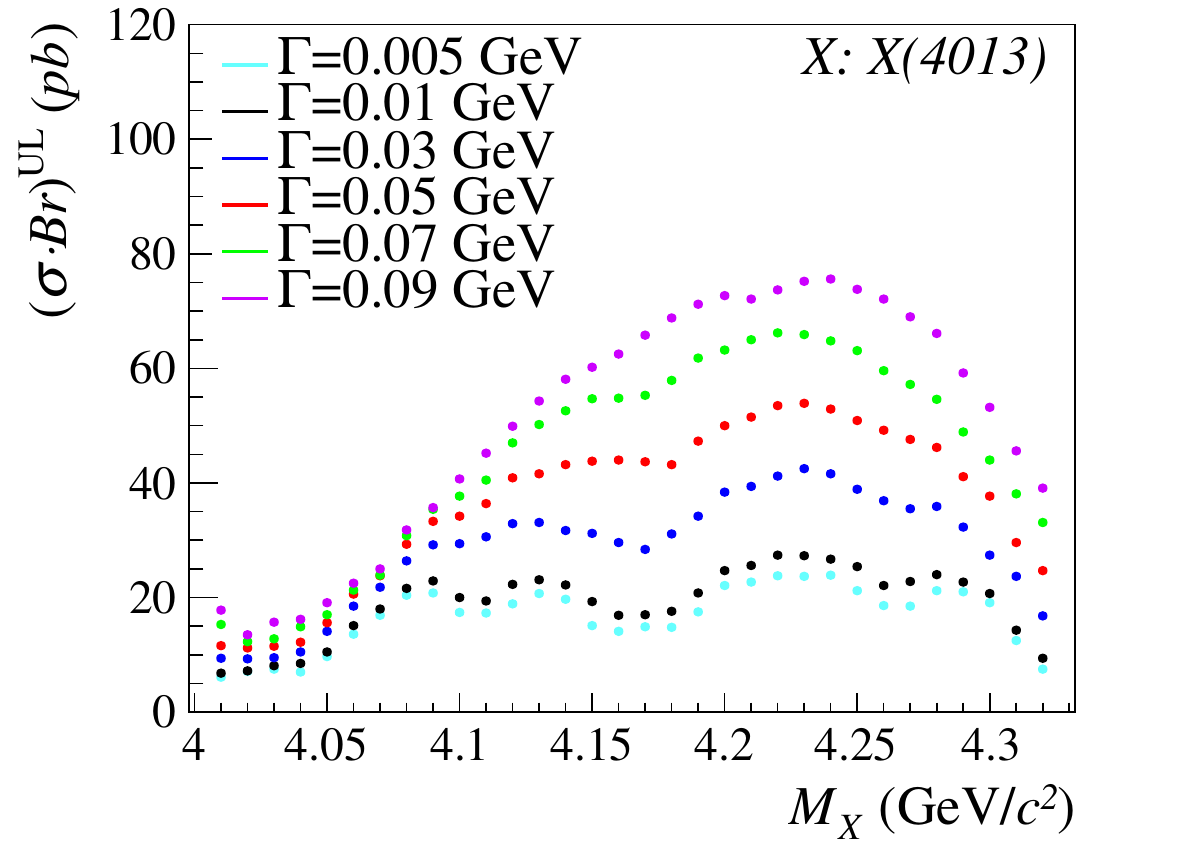}
  \put(50,1.2){(a)}
  \end{overpic}
  \begin{overpic}[width=0.32\textwidth]{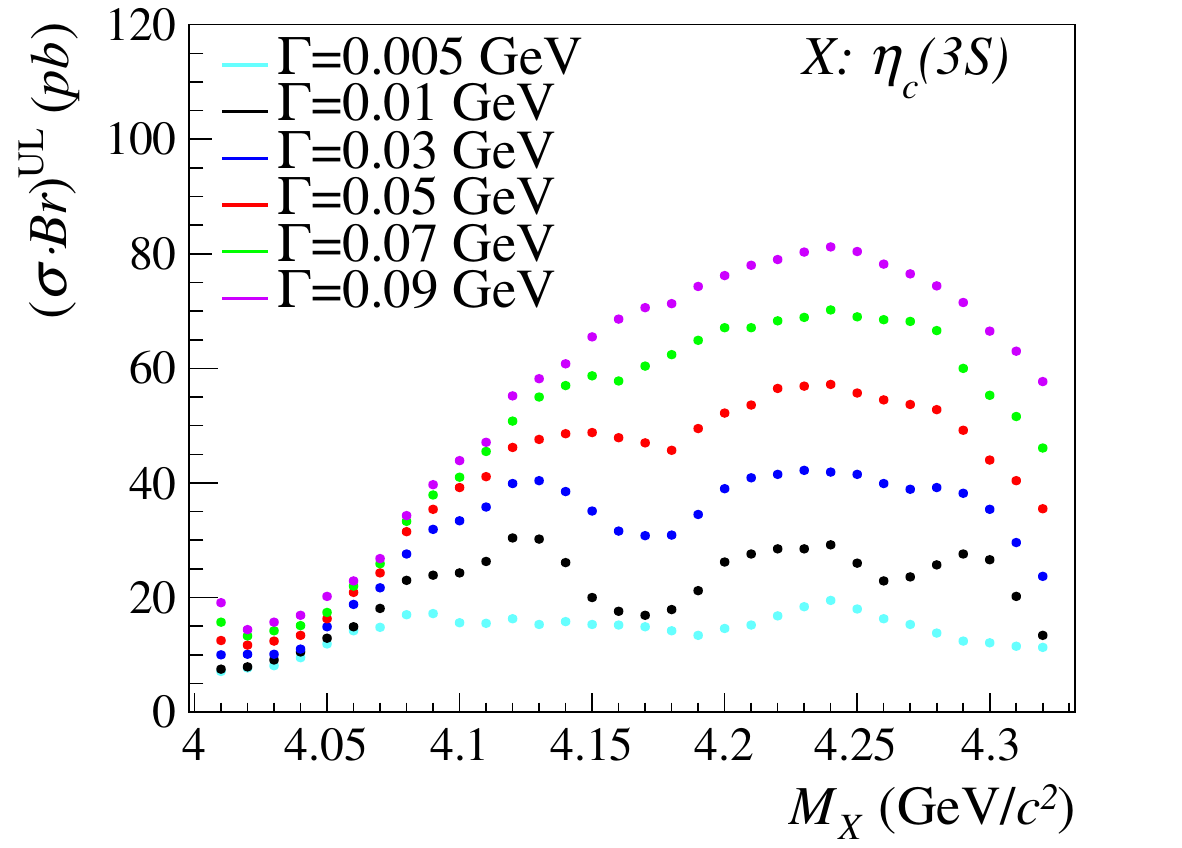}
  \put(50,1.2){(b)}
  \end{overpic}
\centering
  \begin{overpic}[width=0.32\textwidth]{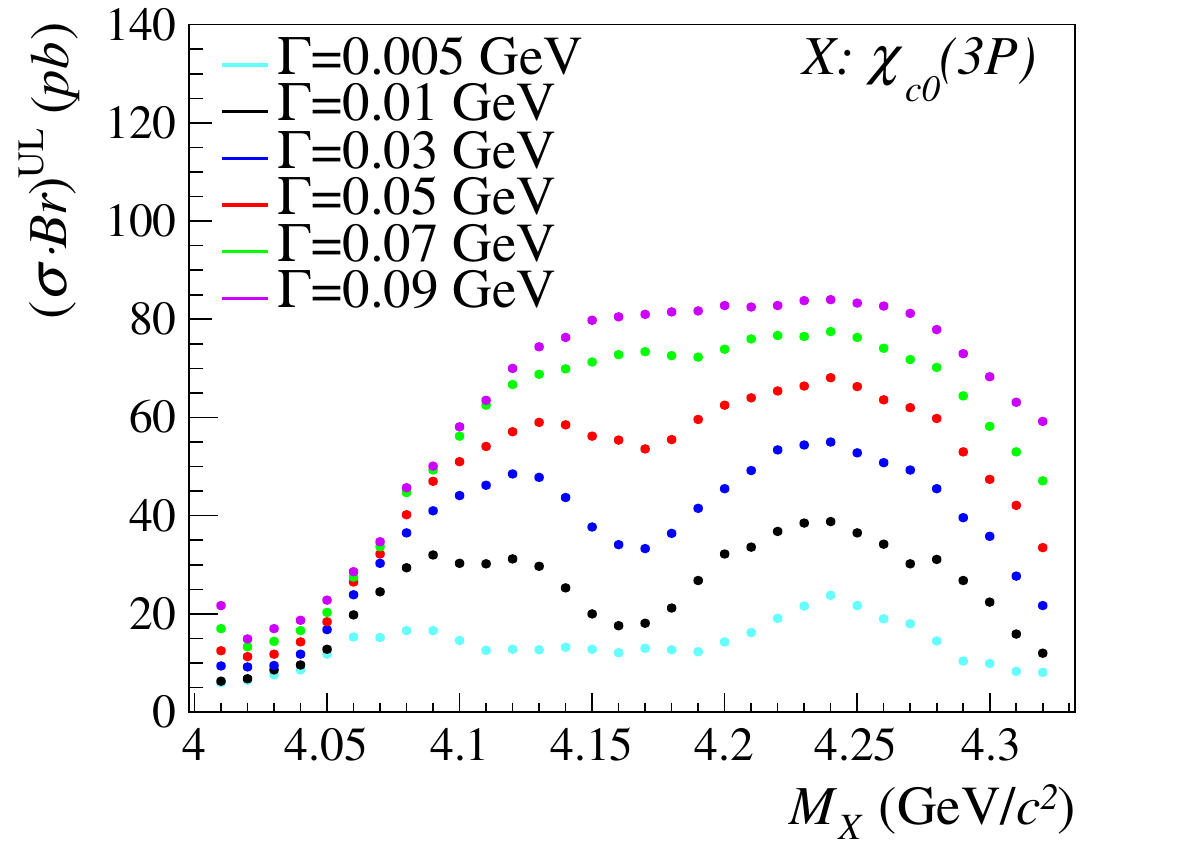}
  \put(50,1.2){(c)}
  \end{overpic}
  \par\vspace{0.15cm}
  \begin{overpic}[width=0.32\textwidth]{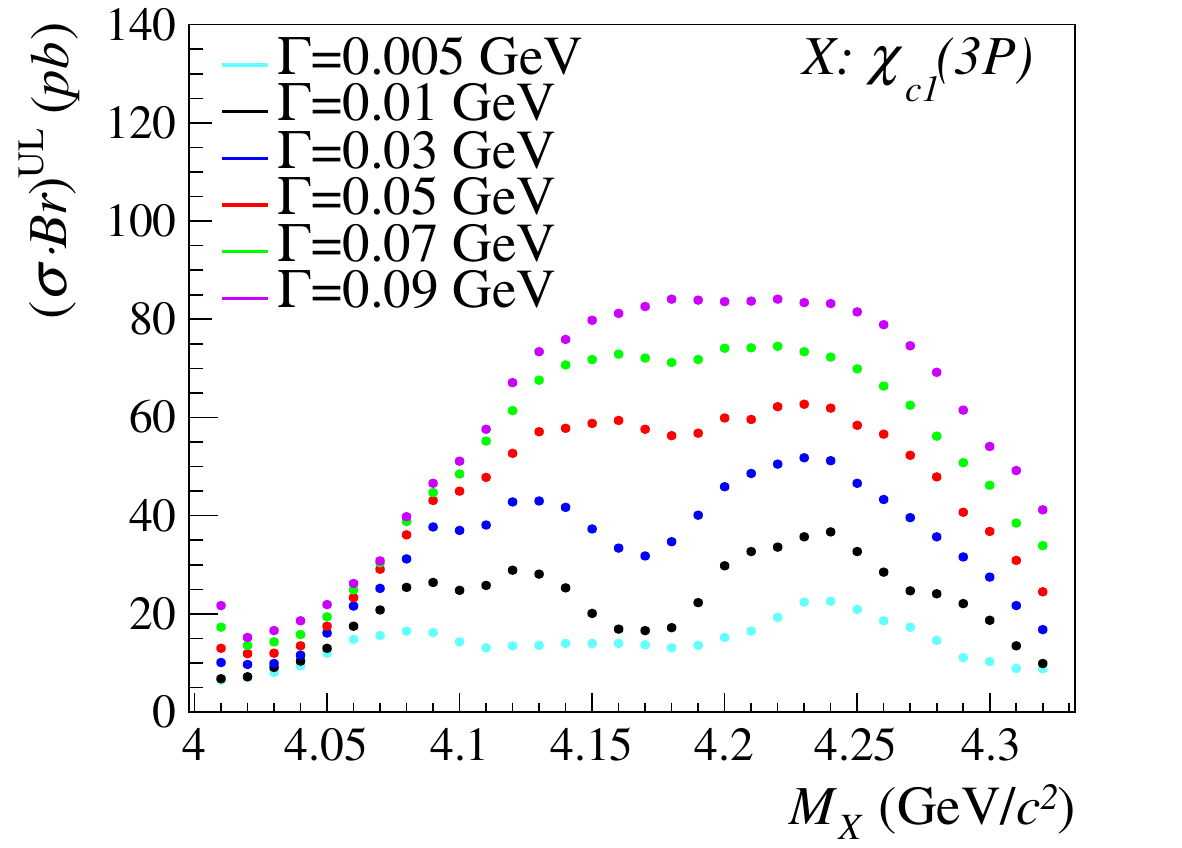}
  \put(50,1.2){(d)}
  \end{overpic}
  \begin{overpic}[width=0.32\textwidth]{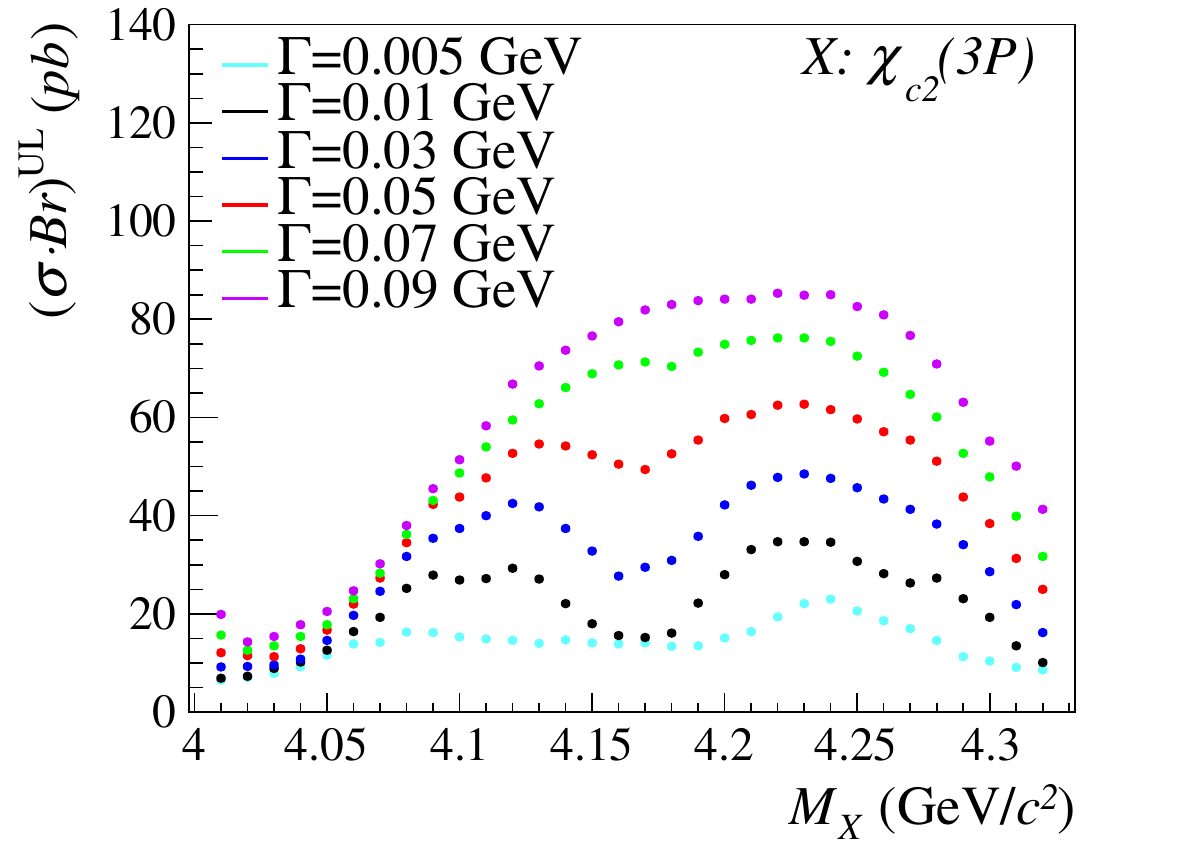}
  \put(50,1,2){(e)}
  \end{overpic}
  
    \caption{The upper limits of the $\sigma\cdot {\rm Br}$ as the function of $X$ masses, after considering all the systematic uncertainties, for different $C$-even states and different widths, under the hypotheses of (a) $X(4013)$, (b) $\eta_c(3S)$, (c) $\chi_{c0}(3P)$, (d) $\chi_{c1}(3P)$ and (e) $\chi_{c2}(3P)$. The cyan, black, blue, red, green, and purple dots are the upper limit distributions when $\Gamma=0.005$, 0.01, 0.03, 0.05, 0.07, and 0.09~GeV, respectively. } 
	\label{different_m_w}
\end{figure*}

\begin{figure*}[htbp] 
\centering
  \begin{overpic}[width=0.32\textwidth]{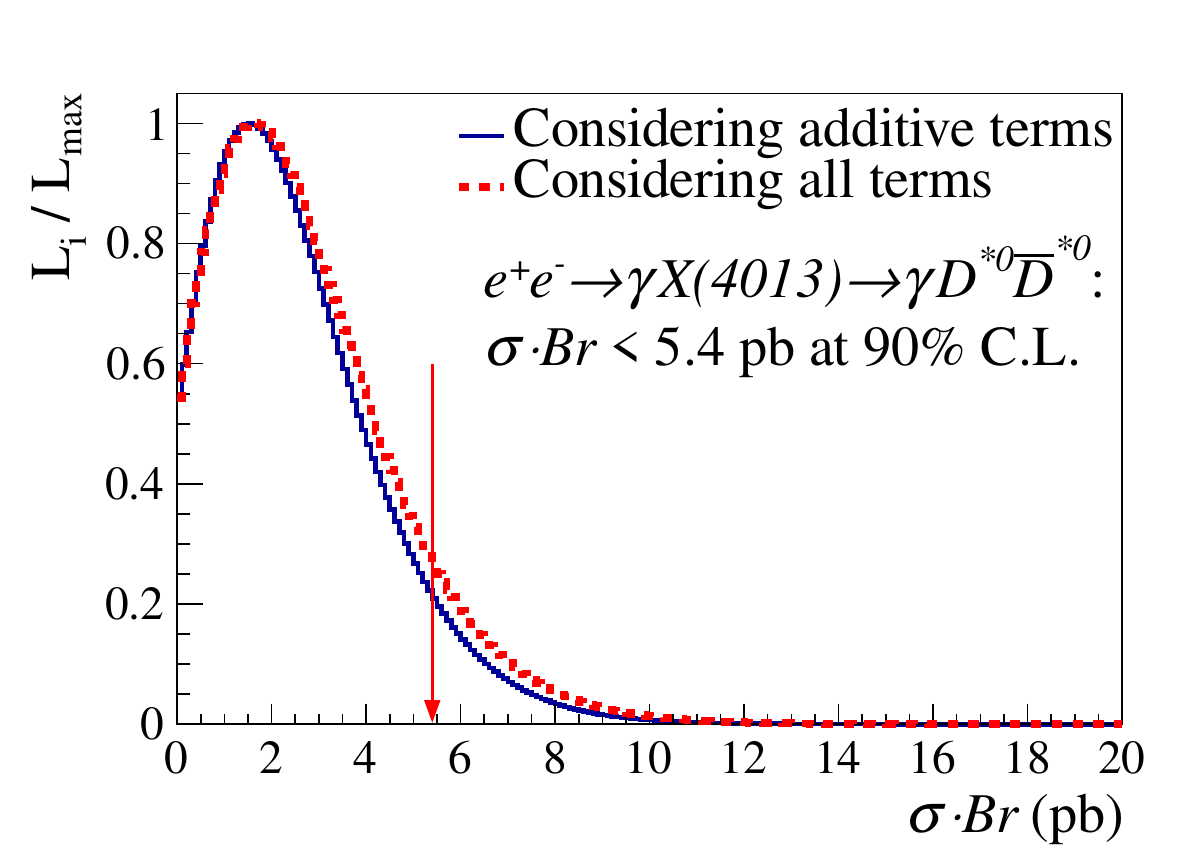}
  \put(50,1.2){(a)}
  \end{overpic}
  \begin{overpic}[width=0.32\textwidth]{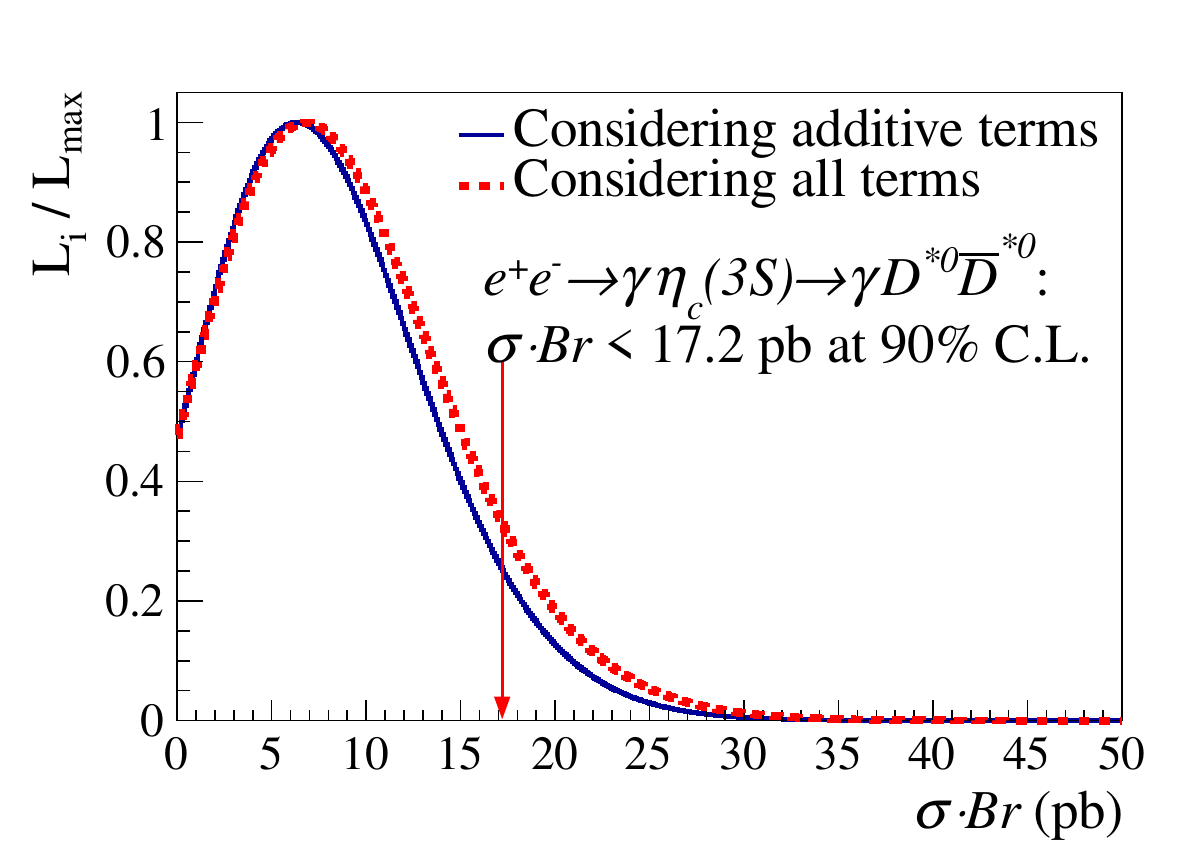}
  \put(50,1.2){(b)}
  \end{overpic}
\centering
  \begin{overpic}[width=0.32\textwidth]{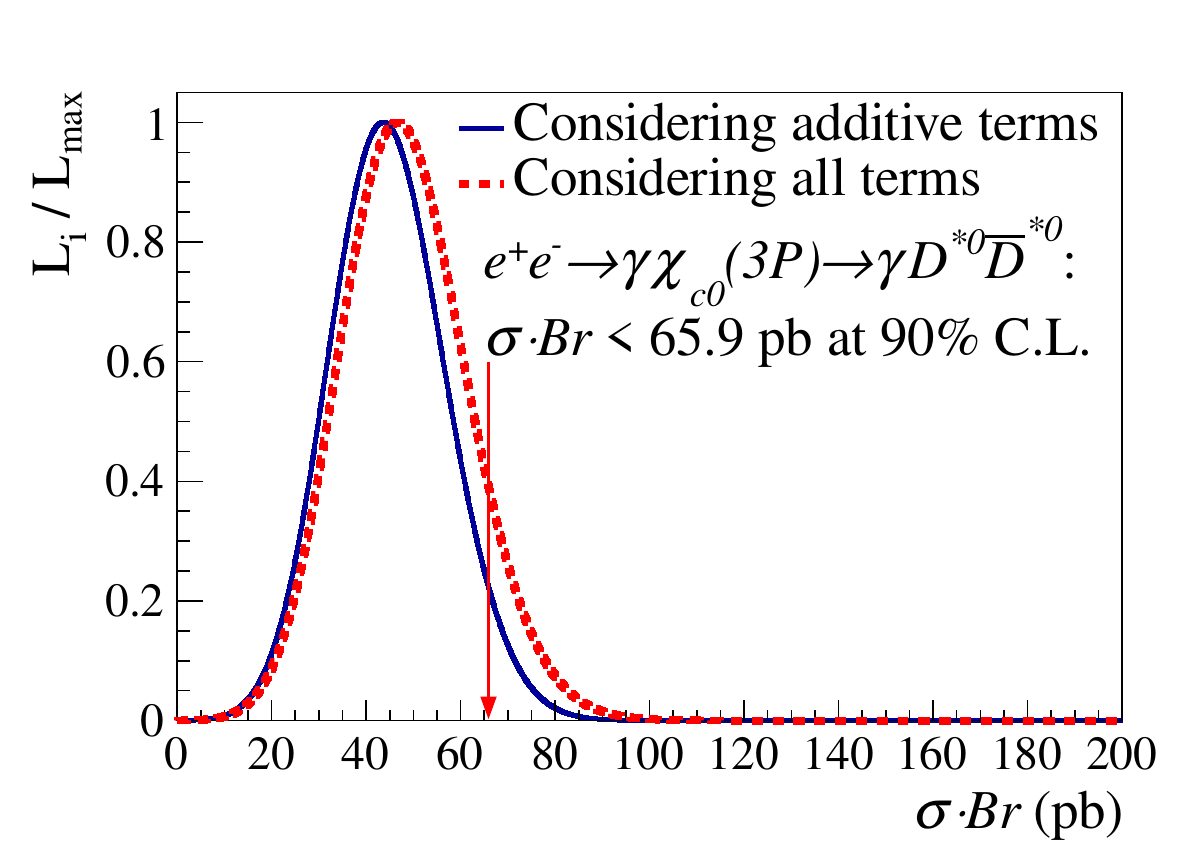}
  \put(50,1.2){(c)}
  \end{overpic}
  \par\vspace{0.15cm}
  \begin{overpic}[width=0.32\textwidth]{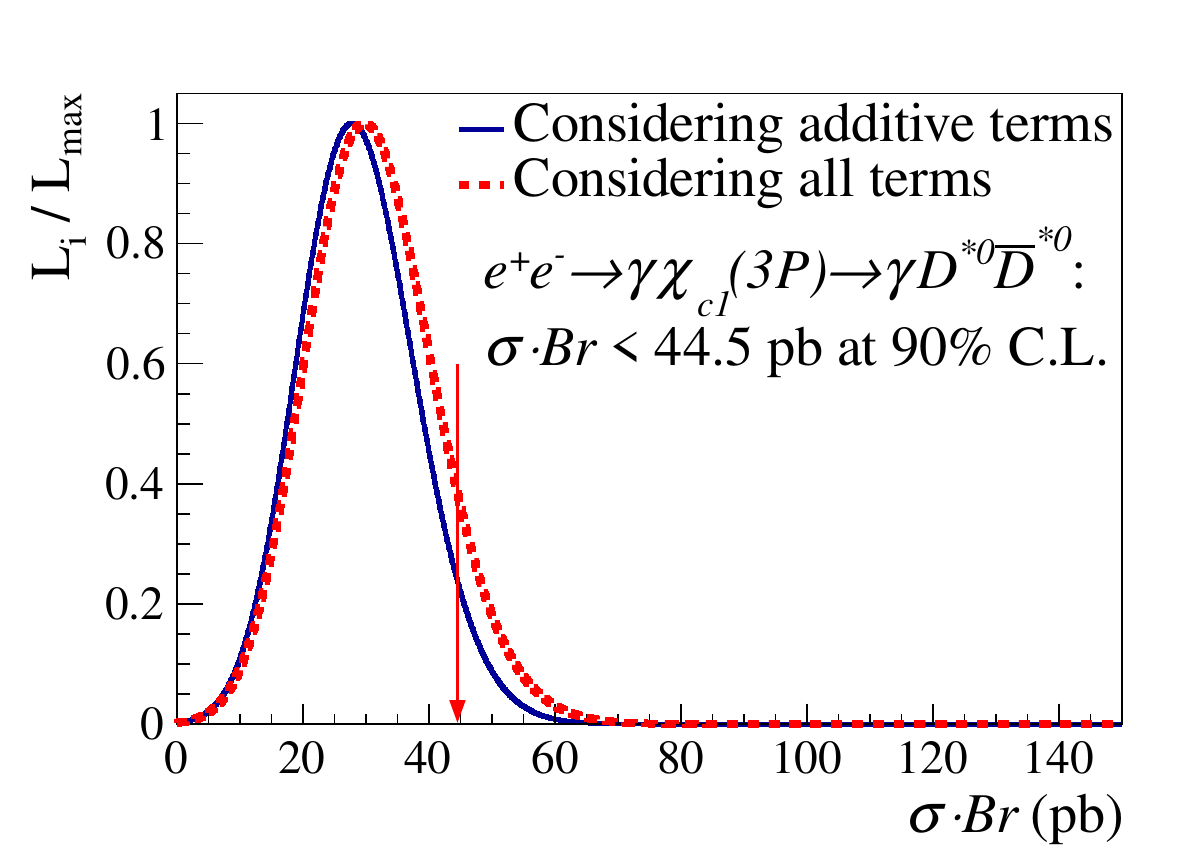}
  \put(50,1.2){(d)}
  \end{overpic}
  \begin{overpic}[width=0.32\textwidth]{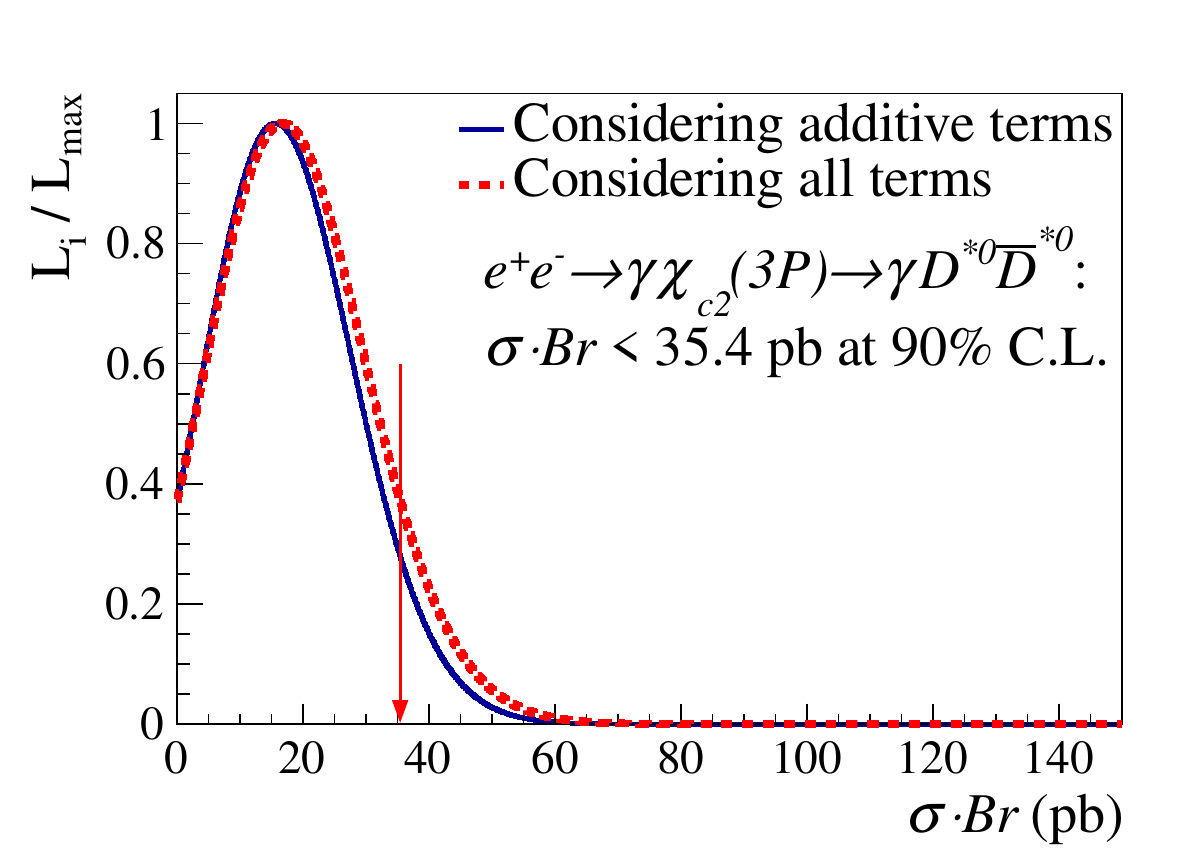}
  \put(50,1.2){(e)}
  \end{overpic}
  
    \caption{The likelihood distributions as a function of the $\sigma\cdot {\rm Br}$ with masses and widths fixed to those of (a) $X(4013)$, (b) $\eta_c(3S)$, (c) $\chi_{c0}(3P)$, (d) $\chi_{c1}(3P)$ and (e) $\chi_{c2}(3P)$ as listed in Table~\ref{five-X}. The solid blue lines show the likelihood distributions considering additive uncertainties, the dashed red lines show the likelihood distributions considering all uncertainties. The red arrows denote the upper limits at the 90\% confidence level.  } 
	\label{upsys}
\end{figure*}

The fit results are shown in Figs.~\ref{simfit}(a, b, c, d, e) for the $\ppDD$ mode and Figs.~\ref{simfit}(f, g, h, i, j) for the $\gampiDD$ mode for $\x, \etac, \chicjp$, respectively. 
The numbers of signal events in the two $\Dsz\Dszb$ decay modes are constrained using the corresponding detection efficiencies and BFs. The result from the simultaneous fit is summarized in Table~\ref{sim}. 
In addition, we perform a $\chi^{2}$ test for the distribution of the fitted curve and data points on the $\Dsz\Dszb$ mass spectrum. The $\chi^{2}/N_{\rm{ndf}}$ is a measure of fit quality, where $N_{\rm{ndf}}$ is the number of degrees of freedom. 
We perform an input and output check based on MC simulated samples to quantify any bias in the fit procedure and find none.

Since no significant signal is observed, and the statistical fluctuations cannot be ignored due to limited statistics, 90\% C.L. upper limits on the $\sigma\cdot {\rm Br}$ are computed, $N_{C.L.(90\%)}$, by solving the
equation~\cite{bayesian}, 
\begin{linenomath*}
\begin{equation}
\int_{0}^{N_{\mathrm{C.L.}}(90\%)} \mathcal{L}(n)\mathrm{d}n \bigg/ \int_{0}^{+\infty} \mathcal{L}(n)\mathrm{d}n = 0.90,
\end{equation} 
\end{linenomath*}
where $n$ is the assumed value of $\sigma\cdot {\rm Br}$, and $\mathcal{L}(n)$ is the corresponding maximized likelihood of the fit with each fixed value of $n$. The resulting 90\% C.L. upper limits on $\sigma\cdot {\rm Br}$, including systematic uncertainties are summarized in Table~\ref{sim}. 

Since the mass and width of the $X$ states under investigation are from theoretical estimations, there are uncertainties from the models. From studies of simulation samples (Group-II), we obtain the upper limits of $\sigma\cdot {\rm Br}$ under different assumptions of the masses and widths of the $X$ states after considering all the systematic uncertainties, as shown in Figs.~\ref{different_m_w}(a,b,c,d,e) for $\x$, $\eta_c(3S)$, $\chi_{c0}(3P)$, $\chi_{c1}(3P)$ and $\chi_{c2}(3P)$, respectively. Detailed descriptions of the systematic uncertainties are provided in the subsequent sections.

\section{SYSTEMATIC UNCERTAINTY}
The systematic uncertainties for $\sigma\cdot {\rm Br}$ are summarized in Table~\ref{sys}. 
They are classified into two categories: multiplicative and additive terms. 

\begin{table*}[htbp] 
\renewcommand{\arraystretch}{1.3}
	\caption{Systematic uncertainties for the $\sigma\cdot {\rm Br}$ measurements [$\%$]. The sources are multiplicative terms.  }
	\begin{center}    
		\begin{tabular}{ c | c | c | c | c | c }  
			\hline
			~~~~Source~~~ &~~~ $\x$   ~~~&~~~ $\etac$ ~~~&~~~ $\chiczp$  ~~~&~~ $\chicop$  ~~~&~~ $\chictp$ ~~\\\hline
			Tracking                   ~&~ 5.8   ~&~ 5.8   ~&~ 5.8   ~&~ 5.8   ~&~ 5.8     \\
			Photon reconstruction      ~&~ 3.5   ~&~ 3.5   ~&~ 3.4   ~&~ 3.4   ~&~ 3.4     \\
            $K_{S}^{0}$ reconstruction      ~&~ 0.2   ~&~ 0.2   ~&~ 0.2   ~&~ 0.2   ~&~ 0.2     \\
            Integrated luminosity      ~&~ 0.5   ~&~ 0.5   ~&~ 0.5   ~&~ 0.5   ~&~ 0.5  \\
            Kinematic fit              ~&~ 0.6   ~&~ 0.9   ~&~ 0.9   ~&~ 0.8   ~&~ 1.2  \\
            ${\rm Br}_{\Dsz\ra\piz/\gamma\Dz,~\piz\ra\gamma\gamma}$   ~&~ 3.3   ~&~ 3.3   ~&~ 3.3   ~&~ 3.3   ~&~ 3.4  \\
			${\rm Br}_{\Dz\ra K^{-}\pi^{+}, K^{-}\pi^{+}\pi^{+}\pi^{-}, \ks\pi^{+}\pi^{-}, \ks\ra\pi^{+}\pi^{-}}$  ~&~ 3.5   ~&~ 3.5   ~&~  3.5  ~&~ 3.5   ~&~ 3.5  \\
			$\Dz$ mass window          ~&~ 0.0   ~&~ 0.0   ~&~ 0.0   ~&~ 0.0   ~&~ 0.0  \\
			$\Dzb$ mass window         ~&~ 0.0   ~&~ 0.0   ~&~ 0.0   ~&~ 0.0   ~&~ 0.0  \\
			$\Dsz$ mass window         ~&~ 0.3   ~&~ 0.3   ~&~ 0.3   ~&~ 0.3   ~&~ 0.3  \\
			$\Dszb$ mass window        ~&~ 0.1   ~&~ 0.1   ~&~ 0.1   ~&~ 0.1   ~&~ 0.1  \\\hline
            Total                      ~&~ 8.4   ~&~ 8.4   ~&~ 8.4   ~&~ 8.4   ~&~ 8.4   \\\hline
   
		\end{tabular}
		
	\end{center}
	\label{sys}
\end{table*}

The multiplicative terms refer to the uncertainties due to the integrated luminosity, the detector 
efficiency, and the BFs. They are estimated separately for the $\ppDD$ and $\gampiDD$ decay modes. 
Varying the corresponding values by $\pm 1\sigma$, the changes in the average detection efficiencies 
are assigned as the systematic uncertainties. They are described in detail in the following.  

The uncertainty from the tracking is assigned to be 1\% per track using the control samples of $\jpsi\ra\piz\pp$ or $\jpsi\ra K_{S}^{0}K^{\pm}\pi^{\mp}$~\cite{sys-track}. 
There are four charged tracks in DC-11, DC-13, DC-31, and DC-33, and six tracks in DC-12, DC-21, DC-23 and DC-32, and eight tracks in DC-22. The charged tracks $(\pi^{+}\pi^{-})$ of the $K_{S}^{0}$ daughters are not counted. The combined uncertainty for nine decay chains is calculated by weighting the branching fraction and the efficiency.
The uncertainty from the photon reconstruction is studied using the control samples $\jpsi\ra\rho^{0}\pi^{0}$ and $e^{+}e^{-}\ra\gamma\gamma$~\cite{sys-photon}, and is assigned to be 1.0\% per photon. 
The uncertainty from the $K_{S}^{0}$ reconstruction is studied using the control samples $\jpsi\ra K^{*}(892)^{\pm}K^{\mp},~K^{*}(892)^{\pm}\ra K_{S}^{0}\pi^{\pm}$ and $\jpsi\ra\phi K_{S}^{0}K^{\pm}\pi^{\mp}$~\cite{sys-ksrec}, and is assigned to be 1.2\% per $K_{S}^{0}$. 
The total uncertainty for the $\ppDD$ and $\gampiDD$ modes is calculated by weighting the branching fraction and the efficiency. The integrated luminosity is measured using Bhabha scattering events with an uncertainty of 0.5\%~\cite{luminosity_4680}. 

In the kinematic fit, the helix parameters of charged tracks in MC samples have been corrected to improve the consistency between data and MC simulations~\cite{helix}. The differences of efficiencies with and without the helix parameter correction are taken as systematic uncertainties. 

The uncertainties from the BFs of $\Dsz\ra\piz\Dz,~\Dsz\ra\gamma\Dz,~\piz\ra\gamma\gamma$ are 1.39\%, 2.55\%, and 0.03\% according the PDG~\cite{pdg}, respectively. The uncertainties from the BFs of $\konepi$, $\kthreepi$, $\kspipi$ and $\ks\ra\pi^{+}\pi^{-}$ are 0.76\%, 1.70\%, 6.43\%, and 0.07\% ~\cite{pdg}, respectively. 

The systematic uncertainties from the $\Dz~(\Dzb)$ and $\Dsz~(\Dszb)$ mass window requirements come from the mass resolution difference between data and MC simulation. This has been studied by using $\ee\ra\Dz\Dszb\ra\Dz\Dzb\piz$
as a control sample. The line-shape of the cross section for $\ee\to\Dz\Dszb$ is from Refs.~\cite{bkg1:1, DDs:1}. The $\Dz$ ($\Dzb$) meson is reconstructed with $\konepi$, $\kthreepi$, and $\kspipi$. The criteria used to select the control sample are similar to that used to select the signal process. 
The mass distributions of $\Dz~(\Dzb)$ and $\Dsz~(\Dszb)$ from data are fitted with the corresponding mass spectra from the simulated shapes convolved with a Gaussian function, respectively. The $\Dz~(\Dzb)$ and $\Dsz~(\Dszb)$ distributions from the control MC samples are smeared with the resultant Gaussian function. The changes of the signal detection efficiencies are taken as the systematic uncertainties. The uncertainties from the $\Dz~(\Dzb)$ mass window requirements are below 0.1\%, and the Barlow test~\cite{barlowtest} shows no obvious trend, allowing them to be ignored.

The additive terms are related to the determination of $N^{\rm sig}$ from the fit. Different fit conditions are tested in the simultaneous fit. In the nominal fit, the line-shapes of $S$ and $B_{1,2,3}$ are extracted from the MC sample obtained by using RooKeysPdf~\cite{keyspdf}. The uncertainty from the corresponding line-shape is estimated by replacing it with RooHistPdf~\cite{histpdf}. As for $B_{4}$, the line-shape is extracted from the remaining inclusive MC sample obtained by using RooKeysPdf~\cite{keyspdf} in the nominal fit. The uncertainty from the corresponding line-shape is estimated by replacing it with an ARGUS function~\cite{ARGUS}. The numbers of $B_{1,2}$ background events are varied by $\pm 1\sigma$ in the simultaneous fit to estimate the systematic uncertainties. 
The smoothing parameter of the RooKeysPdf~\cite{keyspdf} for $B_{1,2,4}$ is set to 0.2 and 0.9 separately to estimate the systematic uncertainty arising from the choice of smoothing level. 

The effect of the systematic uncertainties on the upper limit are considered in two steps. For additive terms, the largest upper limit of $\sigma\cdot {\rm Br}$ from different fit conditions is chosen. The multiplicative systematic uncertainties are incorporated by convolving the likelihood distribution with a Gaussian function~\cite{upper}. The likelihood distributions are shown in Figs.~\ref{upsys}(a,b,c,d,e) for $\x$, $\eta_c(3S)$, $\chi_{c0}(3P)$, $\chi_{c1}(3P)$ and $\chi_{c2}(3P)$, respectively. The upper limit on $\sigma\cdot {\rm Br}$ at 90\% C.L. is determined as summarized in Table~\ref{sim}.

\section{SUMMARY}
Using a sample corresponding to an integrated luminosity of $(1667.4\pm0.2\pm8.8)~\rm{pb^{-1}}$ collected at $\sqrt{s}=4.682$~GeV by the BESIII detector, 
we search for five $C$-even states $X$, decaying into the $\Dsz\Dszb$ pair in the process $\ee\ra\gamma X$ with the two decay modes $\ppDD$ and $\gampiDD$. No significant signals are observed. The upper limits of $\sigma_{\ee\ra\gamma X}\cdot {\rm Br}_{X\ra\Dsz\Dszb}$ at 90\% C.L. are determined to be 5.4 pb, 17.2 pb, 65.9 pb, 44.5 pb, and 35.4 pb, in which $X$ is $\x$, $\etac$, $\chiczp$, $\chicop$, and $\chictp$, respectively, in an assumption of predicted masses and widths. 
The signal yields of $\ee\ra\gamma X(3872)\ra\gamma\Dsz\Dzb$ is $3.80\pm0.68$ pb~\cite{3872-to-dsds}. In this paper, the upper limit for $\ee\ra\gamma\x\ra\gamma\Dsz\Dszb$ is at the same order of magnitude, demonstrating that the current sensitivity has reached a level comparable to known analogous states. This analysis constrains the existence and yield of the partner of $X(3872)$ and offers useful input for future studies about $\Dsz\Dszb$ bound state.


\section*{ACKNOWLEDGMENTS}

The BESIII Collaboration thanks the staff of BEPCII (https://cstr.cn/31109.02.BEPC) and the IHEP computing center for their strong support. This work is supported in part by National Key R\&D Program of China under Contracts Nos. 2025YFA1613900, 2023YFA1606000, 2023YFA1606704; National Natural Science Foundation of China (NSFC) under Contracts Nos. 12375070, 11635010, 11935015, 11935016, 11935018, 12025502, 12035009, 12035013, 12061131003, 12192260, 12192261, 12192262, 12192263, 12192264, 12192265, 12221005, 12225509, 12235017, 12342502, 12361141819; the Chinese Academy of Sciences (CAS) Large-Scale Scientific Facility Program; the Strategic Priority Research Program of Chinese Academy of Sciences under Contract No. XDA0480600; CAS under Contract No. YSBR-101; Shanghai Leading Talent Program of Eastern Talent Plan under Contract No. JLH5913002; 100 Talents Program of CAS; The Institute of Nuclear and Particle Physics (INPAC) and Shanghai Key Laboratory for Particle Physics and Cosmology; ERC under Contract No. 758462; German Research Foundation DFG under Contract No. FOR5327; Istituto Nazionale di Fisica Nucleare, Italy; Knut and Alice Wallenberg Foundation under Contracts Nos. 2021.0174, 2021.0299, 2023.0315; Ministry of Development of Turkey under Contract No. DPT2006K-120470; National Research Foundation of Korea under Contract No. NRF-2022R1A2C1092335; National Science and Technology fund of Mongolia; Polish National Science Centre under Contract No. 2024/53/B/ST2/00975; STFC (United Kingdom); Swedish Research Council under Contract No. 2019.04595; U. S. Department of Energy under Contract No. DE-FG02-05ER41374



\end{document}